\documentclass[twocolumn]{aastex631}

\usepackage{subfigure}
\usepackage[T1]{fontenc} 
\usepackage[utf8]{inputenc} 

\newcommand\tess{TESS}

\newcommand\ms{$\textrm{m~s}^{-1}$}

\newcommand\gcmcubed{$\textrm{g~cm}^{-3}$}

\newcommand\teff{T$_{\rm{eff}}$}
\newcommand\vsini{$v\sin i$}

\newcommand{\unit}[1]{\ensuremath{\, \mathrm{#1}}} 
\newcommand\earthmass{$M_{\oplus}$}
\newcommand\earthradius{$R_{\oplus}$}

\submitjournal{AJ}

\begin{document}

\title{An In-Depth Look at TOI-3884b: a Super-Neptune Transiting a M4 Dwarf with Persistent Star Spot Crossings}

\author[0000-0002-2990-7613]{Jessica E. Libby-Roberts}
\affil{Department of Astronomy \& Astrophysics, 525 Davey Laboratory, The Pennsylvania State University, University Park, PA 16802, USA}
\affil{Center for Exoplanets and Habitable Worlds, 525 Davey Laboratory, The Pennsylvania State University, University Park, PA 16802, USA}

\author[0000-0003-2435-130X]{Maria Schutte}
\affil{Homer L. Dodge Department of Physics and Astronomy, University of Oklahoma, 440 W. Brooks Street, Norman, OK 73019, USA}

\author[0000-0003-1263-8637]{Leslie Hebb}
\affil{Physics Department, Hobart and William Smith Colleges, 300 Pulteney Street, Geneva, NY 14456, USA}
\affil{Department of Astronomy, Cornell University, 245 East Ave, Ithaca, NY 14850, USA}

\author[0000-0001-8401-4300]{Shubham Kanodia}
\affil{Earth and Planets Laboratory, Carnegie Institution for Science, 5241 Broad Branch Road, NW, Washington, DC 20015, USA}
\affil{Department of Astronomy \& Astrophysics, 525 Davey Laboratory, The Pennsylvania State University, University Park, PA 16802, USA}
\affil{Center for Exoplanets and Habitable Worlds, 525 Davey Laboratory, The Pennsylvania State University, University Park, PA 16802, USA}

\author[0000-0003-4835-0619]{Caleb I. Ca\~nas}
\altaffiliation{NASA Postdoctoral Program Fellow}
\affil{NASA Goddard Space Flight Center, 8800 Greenbelt Road, Greenbelt, MD 20771, USA}
\affil{Department of Astronomy \& Astrophysics, 525 Davey Laboratory, The Pennsylvania State University, University Park, PA 16802, USA}
\affil{Center for Exoplanets and Habitable Worlds, 525 Davey Laboratory, The Pennsylvania State University, University Park, PA 16802, USA}

\author[0000-0001-7409-5688]{Guðmundur Stefánsson} 
\affil{NASA Sagan Fellow}
\affil{Department of Astrophysical Sciences, Princeton University, 4 Ivy Lane, Princeton, NJ 08540, USA}

\author[0000-0002-9082-6337]{Andrea S.J.\ Lin}
\affil{Department of Astronomy \& Astrophysics, 525 Davey Laboratory, The Pennsylvania State University, University Park, PA 16802, USA}
\affil{Center for Exoplanets and Habitable Worlds, 525 Davey Laboratory, The Pennsylvania State University, University Park, PA 16802, USA}

\author[0000-0001-9596-7983]{Suvrath Mahadevan}
\affil{Department of Astronomy \& Astrophysics, 525 Davey Laboratory, The Pennsylvania State University, University Park, PA 16802, USA}
\affil{Center for Exoplanets and Habitable Worlds, 525 Davey Laboratory, The Pennsylvania State University, University Park, PA 16802, USA}
\affil{ETH Zurich, Institute for Particle Physics \& Astrophysics, Switzerland}

\author[0000-0001-7142-2997]{Winter Parts(they/them)}
\affil{Department of Astronomy \& Astrophysics, 525 Davey Laboratory, The Pennsylvania State University, University Park, PA 16802, USA}
\affil{Center for Exoplanets and Habitable Worlds, 525 Davey Laboratory, The Pennsylvania State University, University Park, PA 16802, USA}

\author[0000-0002-5300-5353]{Luke Powers}
\affil{Department of Astronomy \& Astrophysics, 525 Davey Laboratory, The Pennsylvania State University, University Park, PA 16802, USA}
\affil{Center for Exoplanets and Habitable Worlds, 525 Davey Laboratory, The Pennsylvania State University, University Park, PA 16802, USA}

\author[0000-0001-9209-1808]{John Wisniewski}
\affil{Department of Physics \& Astronomy, George Mason University, 4400 University Drive, MS 3F3, Fairfax, VA 22030, USA}

\author[0000-0003-4384-7220]{Chad F. Bender}
\affil{Steward Observatory, The University of Arizona, 933 N.\ Cherry Avenue, Tucson, AZ 85721, USA}

\author[0000-0001-9662-3496]{William D. Cochran}
\affil{McDonald Observatory and Department of Astronomy, The University of Texas at Austin, USA}
\affil{Center for Planetary Systems Habitability, The University of Texas at Austin, USA}

\author[0000-0002-2144-0764]{Scott A. Diddams}
\affil{Electrical, Computer \& Energy Engineering, University of Colorado, 425 UCB, Boulder, CO 80309, USA}
\affil{Department of Physics, University of Colorado, 2000 Colorado Avenue, Boulder, CO 80309, USA}
\affil{Time and Frequency Division, National Institute of Standards and Technology, 325 Broadway, Boulder, CO 80305, USA}

\author[0000-0002-0885-7215]{Mark~E.~Everett}
\affiliation{NSF’s National Optical-Infrared Astronomy Research Laboratory, 950 N. Cherry Ave., Tucson, AZ 85719, USA}

\author[0000-0002-5463-9980]{Arvind F.\ Gupta}
\affil{Department of Astronomy \& Astrophysics, 525 Davey Laboratory, The Pennsylvania State University, University Park, PA 16802, USA}
\affil{Center for Exoplanets and Habitable Worlds, 525 Davey Laboratory, The Pennsylvania State University, University Park, PA 16802, USA}

\author[0000-0003-1312-9391]{Samuel Halverson}
\affiliation{Jet Propulsion Laboratory, California Institute of Technology, 4800 Oak Grove Drive, Pasadena, CA 91109, USA
}

\author[0000-0002-4475-4176]{Henry A. Kobulnicky}
\affil{Department of Physics \& Astronomy, University of Wyoming, Laramie, WY 82070, USA}

\author[0000-0001-7458-1176]{Adam F. Kowalski}
\affiliation{National Solar Observatory, University of Colorado Boulder, 3665 Discovery Drive, Boulder, CO 80303, USA}
\affiliation{Department of Astrophysical and Planetary Sciences, University of Colorado, Boulder, 2000 Colorado Ave, CO 80305, USA}
\affiliation{Laboratory for Atmospheric and Space Physics, University of Colorado Boulder, 3665 Discovery Drive, Boulder, CO 80303, USA.}

\author[0000-0002-2401-8411]{Alexander Larsen}
\affil{Department of Physics \& Astronomy, University of Wyoming, Laramie, WY 82070, USA}

\author[0000-0002-0048-2586]{Andrew Monson}
\affil{Steward Observatory, The University of Arizona, 933 N.\ Cherry Avenue, Tucson, AZ 85721, USA}

\author[0000-0001-8720-5612]{Joe P.\ Ninan}
\affil{Department of Astronomy and Astrophysics, Tata Institute of Fundamental Research, Homi Bhabha Road, Colaba, Mumbai 400005, India}

\author[0000-0001-9307-8170]{Brock A. Parker}
\affil{Department of Physics \& Astronomy, University of Wyoming, Laramie, WY 82070, USA}

\author[0000-0002-4289-7958]{Lawrence W. Ramsey}
\affil{Department of Astronomy \& Astrophysics, 525 Davey Laboratory, The Pennsylvania State University, University Park, PA 16802, USA}
\affil{Center for Exoplanets and Habitable Worlds, 525 Davey Laboratory, The Pennsylvania State University, University Park, PA 16802, USA}

\author[0000-0003-0149-9678]{Paul Robertson}
\affil{Department of Physics \& Astronomy, University of California Irvine, Irvine, CA 92697, USA}

\author[0000-0002-4046-987X]{Christian Schwab}
\affil{School of Mathematical and Physical Sciences, Macquarie University, Balaclava Road, North Ryde, NSW 2109, Australia}

\author[0000-0002-5817-202X]{Tera N. Swaby}
\affil{Department of Physics \& Astronomy, University of Wyoming, Laramie, WY 82070, USA}

\author[0000-0002-4788-8858]{Ryan C. Terrien}
\affil{Carleton College, One North College St., Northfield, MN 55057, USA}

\correspondingauthor{Jessica E. Libby-Roberts}
\email{jer5346@psu.edu}

\begin{abstract}

We perform an in-depth analysis of the recently validated TOI-3884 system, an M4 dwarf star with a transiting super-Neptune. Using high precision light curves obtained with the 3.5 m Apache Point Observatory and radial velocity observations with the Habitable-zone Planet Finder (HPF), we derive a planetary mass of 32.6$^{+7.3}_{-7.4}$ M$_{\oplus}$ and radius of 6.4 $\pm$ 0.2 R$_{\oplus}$. We detect a distinct star spot crossing event occurring just after ingress and spanning half the transit for every transit. We determine this spot feature to be wavelength-dependent with the amplitude and duration evolving slightly over time. Best-fit star spot models show that TOI-3884b possesses a misaligned ($\lambda$ = 75 $\pm$ 10$^\circ$) orbit which crosses a giant pole-spot. This system presents a rare opportunity for studies into the nature of both a misaligned super-Neptune and spot evolution on an active mid-M dwarf.

\end{abstract}


\section{Introduction} \label{sec:intro}

Giant planets larger than 6 R$_\oplus$ are notably infrequent around FGK-dwarf stars compared to smaller sub-Neptunes and super-Earths \citep{howard_planet_2012}. Giant planets orbiting M dwarfs are even rarer with $<$ 15 discovered to date \citep[e.g.,][]{canas_toi1899_2020,kanodia_toi-532b_2021,jordan_hats-74ab_2021,canas_two_2022}. This sparsity was predicted by \citet{laughlin.core.accretion} who postulated the smaller protoplanetary disks should make it near-impossible for cores to accrete and experience runaway growth within the disks' lifetimes. \textit{TESS} \citep{ricker.tess}, however, continues to discover new giant planets orbiting M dwarfs. All previously discovered giant planets orbit early- and mid-M dwarfs with stellar masses $>$ 0.35 M$_\odot$ \citep{kanodia_5205_2022}. 

TOI-3884b is the first transiting super-Neptune discovered orbiting a M4 Dwarf with a stellar mass of 0.30 M$_\odot$. Its planetary nature was originally validated by \citet{almenara.toi3884}, who obtained several ground-based transits with ExTrA \citep{extra.instrument} and LCOGT \citep{brown.lcogt} as well as two radial velocity (RV) points with ESPRESSO \citep{pepe.espresso}. Interestingly, TOI-3884b possesses a persistent signature in every transit indicative of a star spot crossing event. Given the lack of notable out-of-transit variability, \citet{almenara.toi3884} suggest the spot is a long-lived pole-spot. 

Pole-spots are a common feature on young M dwarfs like TOI-3884. These spots can persist beyond 6-12 months \citep[e.g.,][]{davenport_spotlifetime_2015,robertson_persistent_2020}. In-transit spot crossing events provide an interesting probe for monitoring spot evolution \citep{sanchis_ojeda_hatp11_2011,schutte_starspot_2022}. As the planet passes over a cooler and darker spot, the amount of flux blocked by the planet decreases yielding a bump in the transit light curve \citep[e.g.,][]{sanchis_ojeda_hatp11_2011,morris2017,schutte_starspot_2022}. For TOI-3884b, \citet{almenara.toi3884} used the duration and wavelength-dependent amplitude of this feature to approximate the spot temperature and area. Assuming a polar location, they also estimated the orbital obliquity concluding that TOI-3884b must be misaligned relative to its star's spin-orbit axis. 

The TOI-3884 system is a promising target for future JWST observations. TOI-3884b possesses the highest transmission spectroscopy signal-to-noise ratio per transit for a planet with an equilibrium temperature $<$ 500 K making it a favorable planet for atmospheric characterization. With the assured spot crossing, the transit of TOI-3884b may also provide a direct measure of the spot's impact on the atmospheric transmission spectrum of the planet \citep{rackham_tlse_2018}. 

In this paper, we present an in-depth analysis of the TOI-3884 system. We describe our observations in Section~\ref{sec:observations} which we use to derive updated stellar and planetary parameters in Section~\ref{sec:analysis}. We perform a detailed analysis of the stellar spots in Section~\ref{sec:starspot}. Section~\ref{sec:discussion} discusses these results, as well as places TOI-3884b in context to the growing M dwarf giant planet population. We conclude in Section~\ref{sec:conclusion}.

\section{Observations and Data Reduction}\label{sec:observations}

\subsection{TESS}\label{sec:TESS}

TOI-3884 (TIC 86263325; T$_{\rm{mag}}$ = 12.91; J$_{\rm{mag}}$ = 11.13)\footnote{https://exofop.ipac.caltech.edu} was flagged as an object of interest host in the \textit{TESS} Sector 22 (2020 February 19 -- 2020 March 17) long cadence (30-minute) data by the TESS Quick Look Pipeline \citep[QLP;][]{huang_photometry_2020} during the \textit{Faint-Star Search} \citep{kunimoto.qlp}. The transit shape was noted to show an unusual shape by the TESS Follow-up Observing Program (TFOP)\footnote{https://tess.mit.edu/followup/}. TOI-3884 was again observed by \textit{TESS} in Sector 46 (2021 December 04 -- 2021 December 30) and Sector 49 (2022 March 01 -- 2022 March 25) with 2-minute exposures. 

\begin{figure*}
  \centering
\begin{subfigure}{}  
 \includegraphics[width=\textwidth]{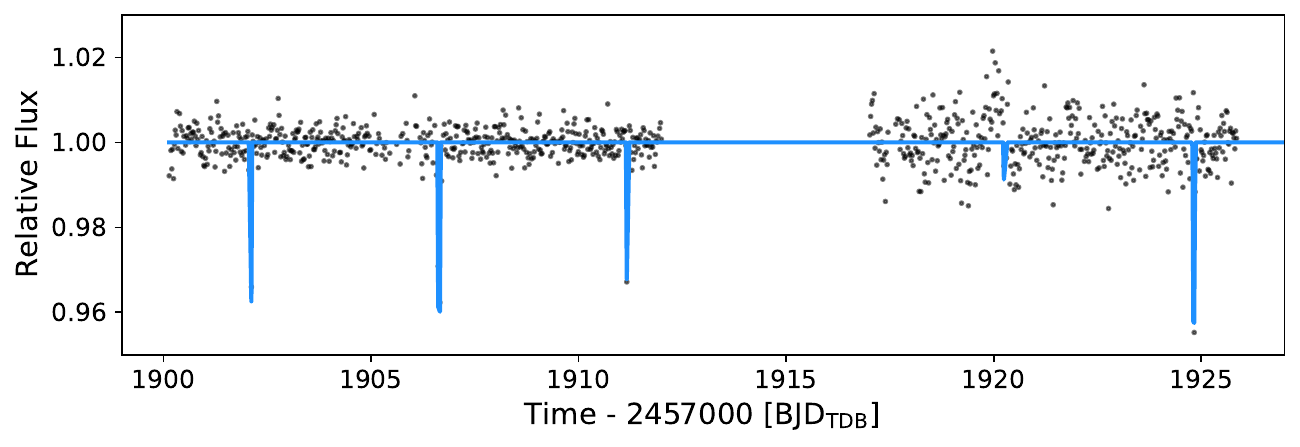}
\end{subfigure}

\begin{subfigure}{}
 \includegraphics[width=\textwidth]{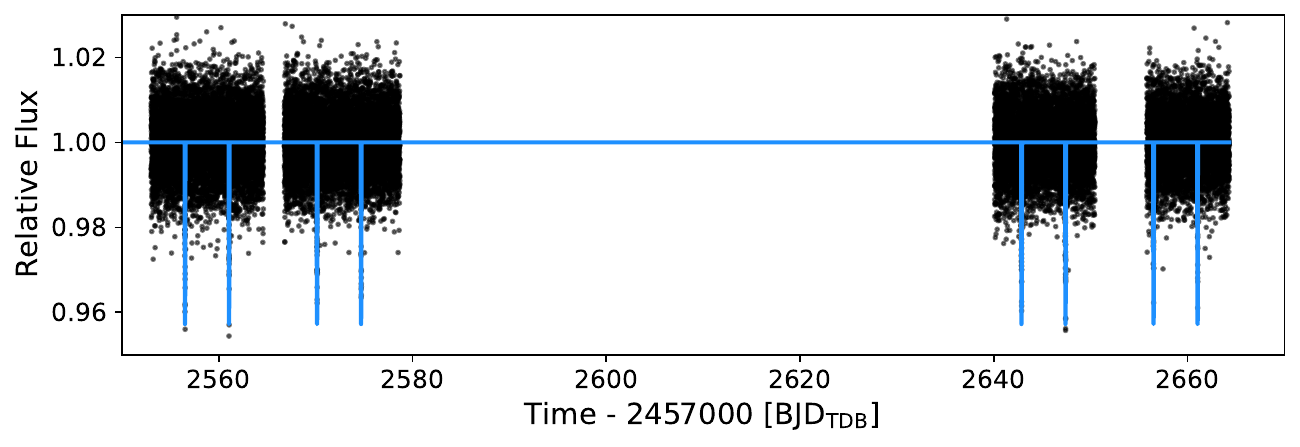}
\end{subfigure}
\caption{\textbf{Top:} \textit{TESS} Sector 22 long-cadence light curve with the TOI-3884b transits denoted in blue. Differing transit depths is an artifact of the 30-minute cadence. \textbf{Bottom:} Short 2-minute cadence of the \textit{TESS} Sectors 46 and 49 with the transit model in blue. Both sets of light curves use the PDCSAP flux without additional out-of-transit GP detrending required.}
\label{fig:LCs}
\end{figure*}

We used the \texttt{lightkurve} package \citep{lightkurve_collaboration_lightkurve_2018} to download all three sectors assuming a `harder' quality flag, removing all NaNs and initial outliers from the Pre-search Data Conditioning Simple Aperture Photometry \citep[PDCSAP; Figure~\ref{fig:LCs}][]{jenkins.spoc,spoc.pipeline}. Folding the 2-minute short cadence light curves in both Sectors 46 and 49 on the expected 4.56 day period for TOI-3884b clearly shows an unusual transit shape (Figure~\ref{fig:full_LCs}) -- an ingress, a bump that spans half the transit, and then the continuation of a normal transit shape through egress. This bump is also present in Sector 22 though the long 30-minute cadence is too sparse to resolve any structure in-transit.

\subsection{Ground-Based Transit Photometric Follow-up}

\begin{figure*}
   \centering
    \includegraphics[width=\textwidth]{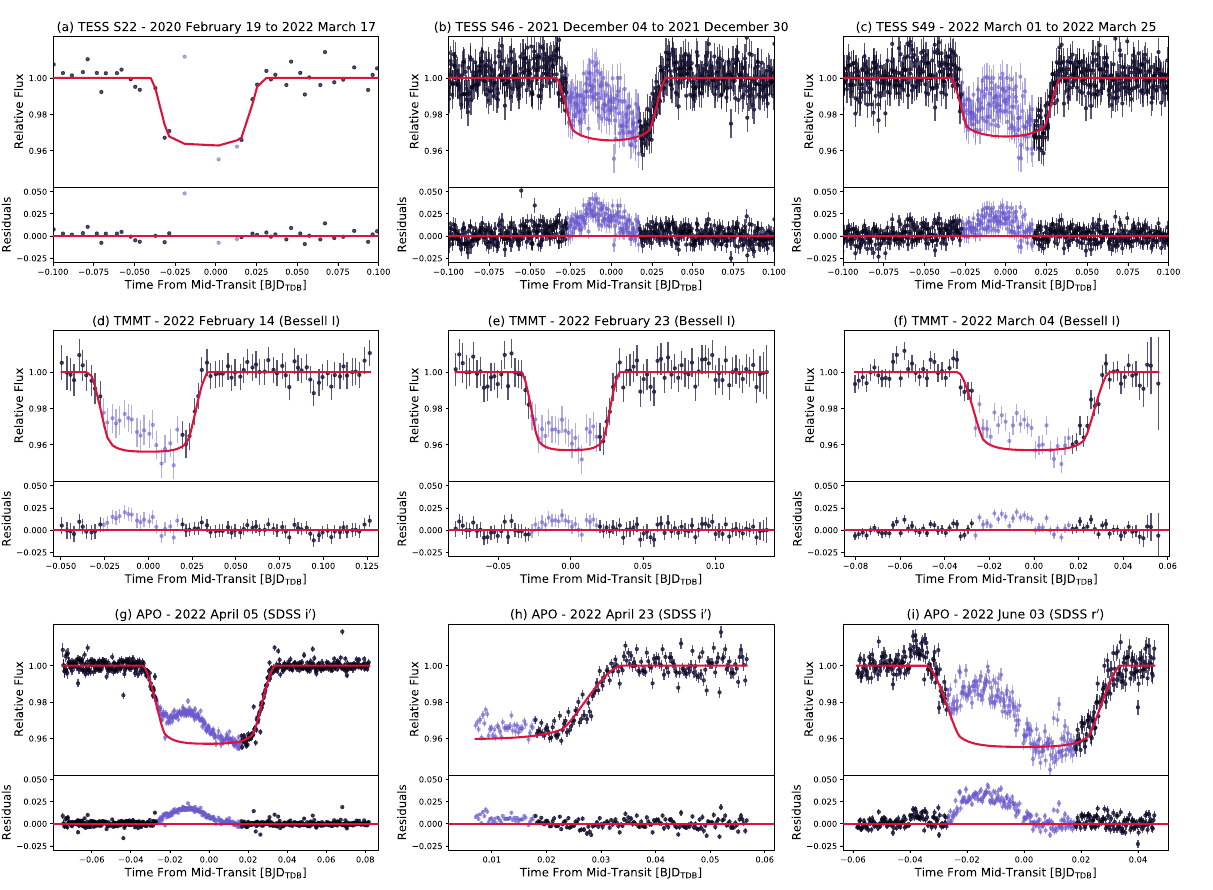}
    \caption{Light curves for individual ground-based observations and the phase-folded \textit{TESS} Sectors 22, 46, and 49. Light blue points were masked in order to fit the transit shape during our analysis and the best-fit non-spotted transit model is plotted in red with the appropriate dilution terms included for the \textit{TESS} sectors (0.98, 0.86, 0.84 respectively). Residuals for the respective transit models are plotted in the bottom panel for each light curve.}
    \label{fig:full_LCs}
\end{figure*}

We observed seven photometric transits/partial-transits of TOI-3884b using three separate ground-based facilities using Bessell I, SDSS i$^\prime$, and SDSS r$^\prime$ filters. We highlight each set of observations below and plot each individual transit, along with the folded TESS transits for Sectors 22, 46, and 49 in Figure~\ref{fig:full_LCs}.

\subsubsection{0.3 m TMMT}
We observed three separate transits of TOI-3884b (2022 February 14, 2022 February 23, 2022 March 4 UT) using the 0.3 m Three-hundred MilliMeter Telescope \citep[TMMT;][]{monson_standard_2017} at Las Campanas Observatory in Chile. Each night used the Bessell I filter with 180 second exposures. Every observation included the entire transit though pre- and post-transit baselines did not span the same length of time. Images collected during each night were then reduced following the procedure highlighted in \citet{monson_standard_2017}. 

We perform aperture photometry on the reduced TMMT images using \texttt{AstroImageJ} \citep{collins_astroimagej}. We assume a photometric aperture radius of 10 pixels (11.9 arcseconds) around the target and 14 reference stars while the median background value was derived from an annulus with inner and outer radii of 15 pixels (17.9 arcseconds) and 25 pixels (29.5 arcseconds) respectively before being subtracted. We divided the target star's flux by the combined flux from the reference stars and derived the flux uncertainties from a combination of stellar, background, and dark current photon noise plus the expected read noise of the instrument. We detrend the light curves by dividing out a linear out-of-transit best-fit model. A similar bump in the transit light curve was present in all three observations.

\subsubsection{3.5 m ARC Telescope}
We observed two transits of TOI-3884b on 2022 April 05 and 2022 June 03 and a partial transit on 2022 April 23 with the ARC 3.5 m Telescope at the Apache Point Observatory (APO) in New Mexico. For all three nights we used the optical CCD Camera ARCTIC equipped with an engineered diffuser \citep{stefansson_toward_2017}. As discussed in \citet{stefansson_toward_2017}, the diffuser enables near photon/scintillation-limited precision light curves by spreading the stellar PSF into a stable top-hat profile without the need to defocus the telescope.

The observations for each night applied the same instrument set-up: quad and fast read-out mode, 4$\times$4 pixel binning, and 20-second exposures. Biases and dome flats were collected either before or after each observing run. ARCTIC does not experience significant dark current for exposures $<$ 60 seconds and was not accounted for in our reduction.

On 2022 April 05 we observed the full transit using the SDSS i$^\prime$ filter with good weather and photometric skies. We also used the SDSS i$^\prime$ filter for the 2022 April 23 transit, though poor weather caused us to miss the first half of the transit and led to significant scatter in the data. To check for chromaticity both in the bump and in the overall transit depth, we observed TOI-3884b on 2022 June 03 using the SDSS r$^\prime$ filter. We experienced non-photometric skies due to dusty conditions.

We reduce each observation with bias subtraction before dividing by a nightly median combined normalized flat field. Aperture photometry was again applied using \texttt{AstroImageJ} assuming an aperture size of 20 pixels (9.1 arcseconds), 5 reference stars, and background annulus of 25 (11.4 arcseconds) and 30 (14.7 arcseconds) pixels for inner and outer radii. Similar to TMMT, we detrend the data by dividing out a linear model calculated from the out-of-transit points. On 2022 June 03, we observed a slight increase in flux prior to transit beyond the linear model which we attributed to a potential micro-flare.

\subsubsection{0.6 m RBO}

We observed the 2022 June 03 transit ingress using the Bessell I filter with the 0.6 m telescope at the Red Buttes Observatory (RBO) in Wyoming, though weather created significant scatter in the transit. While we opted not to include this transit in the analysis, we observed the same slight increase in flux prior to transit as the 2022 June 03 transit obtained with APO. This confirmed the feature is astrophysical and not instrumental or weather-related.

\subsection{NESSI High Contrast Imaging}
We exclude potential background sources that may impact the overall transit signal (depth or shape) using the NN-EXPLORE Exoplanet Stellar Speckle Imager \citep[NESSI;][]{scott_nn-explore_2018} on the WIYN 3.5 m telescope at Kitt Peak National Observatory (KPNO) in Arizona\footnote{The WIYN Observatory is a joint facility of the NSF's National Optical-Infrared Astronomy Research Laboratory, Indiana University, the University of Wisconsin-Madison, Pennsylvania State University, the University of Missouri, the University of California-Irvine, and Purdue University.}. We took a 9-minute sequence of 40 ms exposures using NESSI$^\prime$s z$^{\prime}$ filter on 2022 April 18. These images were then combined and processed following the methods highlighted in \citet{howell_speckle_2011}.

\begin{figure}[]
    \centering
    \includegraphics[width=0.5\textwidth]{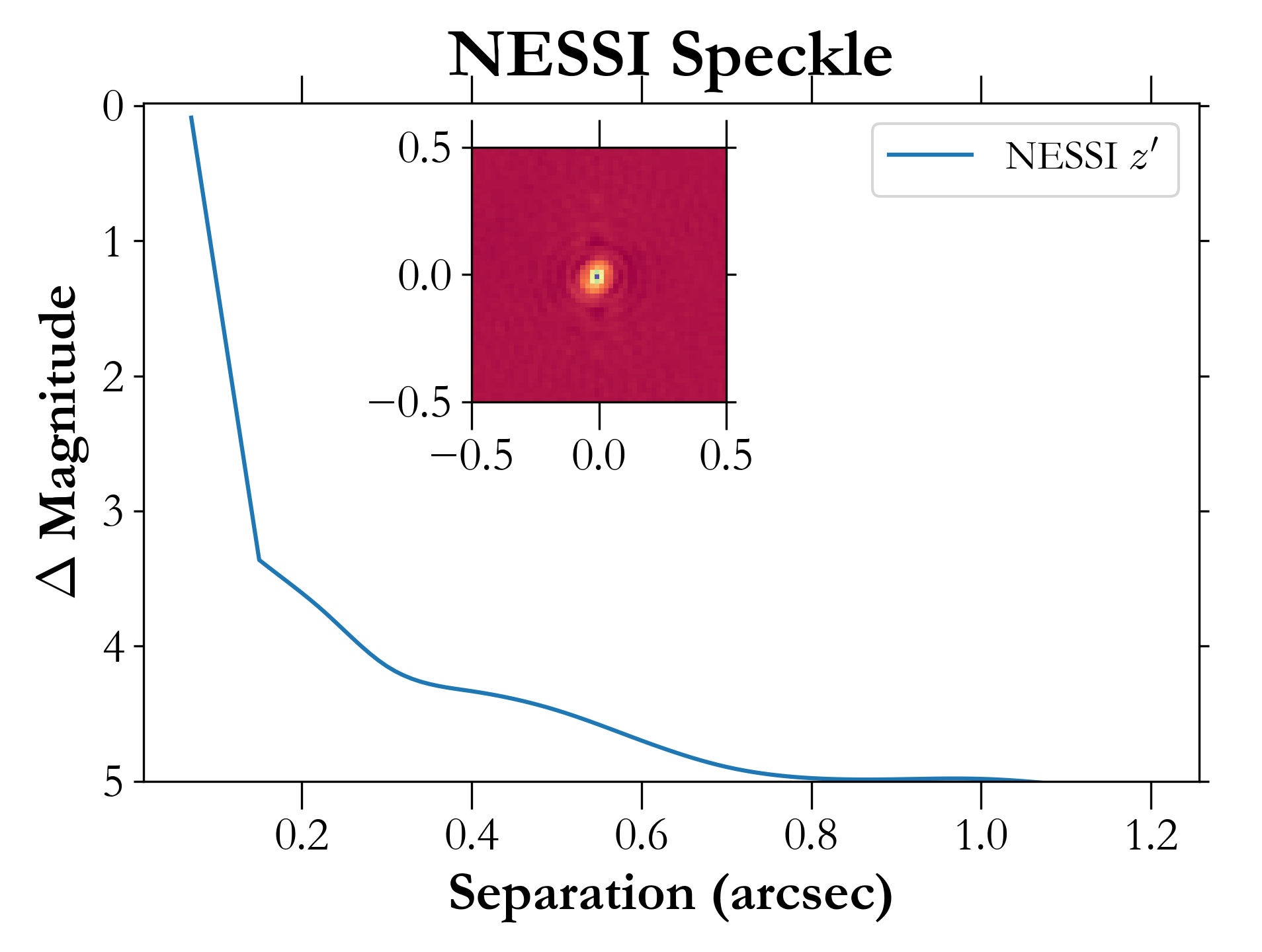}
    \caption{NESSI 5$\sigma$ contrast curve of TOI-3884 with the \(z^\prime\) filter. The inserted image is the final speckle image which shows no nearby sources with $\Delta\rm{mag} > 3.5$ outside 0.2 arcsec}
    \label{fig:NESSI}
\end{figure}

We plot the final contrast curve and speckle image in Figure~\ref{fig:NESSI}. We detect no nearby sources with a $\Delta$z$^{\prime}$ magnitude brighter than 3.8 from 0.2 out to 0.8 arcseconds and magnitudes brighter than 5 from 0.8 out to 1.2 arcseconds. We compliment this with archival Gaia DR3 Data \citep{gaia.dr3.release} which finds no nearby sources within 20 arcseconds. Gaia also assigns TOI-3884 a Renormalized Unit Weight Error (RUWE) equal to 1.25 which is consistent with a single star \citep{ziegler_ruwe_2020,gaia.ruwe}. TOI-3884 is a single star in a fairly sparse region of the night sky.

\subsection{HPF Radial Velocity Follow-Up}
We performed an intensive RV follow-up campaign of TOI-3884 using Habitable-zone Planet Finder \citep[HPF][]{mahadevan_habitable-zone_2012,mahadevan_habitable-zone_2014} starting on 2021 December 01. HPF is a high resolution (R $\sim$55,000) near-infrared (810 -- 1280 nm), fiber-fed \citep{kanodia_overview_2018}, stabilized \citep{stefansson.et.al.2016} precision RV spectrograph, on the 10 m Hobby-Eberly Telescope in Texas \citep{ramsey.het}. Over the next 5 months, we observed TOI-3884 on 27 nights with each night obtaining two 945-second exposure measurements. Each spectrum was analyzed using the \texttt{HxRGproc} package which corrects for bias, non-linearities, cosmic rays, and then calculates the flux and variance of the individual spectra as described in \citet{ninan_habitable-zone_2018}. We use \texttt{barycorrpy} \citep{kanodia_python_2018} to perform the barycentric correction on the individual spectra, which is the Python implementation  of the algorithms from \cite{wright_barycentric_2014}.  A wavelength solution was created by interpolating the wavelength over all other exposures in the same night of each observation, which was then applied to the respective TOI-3884 spectra.

\begin{figure*}[]
    \centering
    \includegraphics[width=\linewidth]{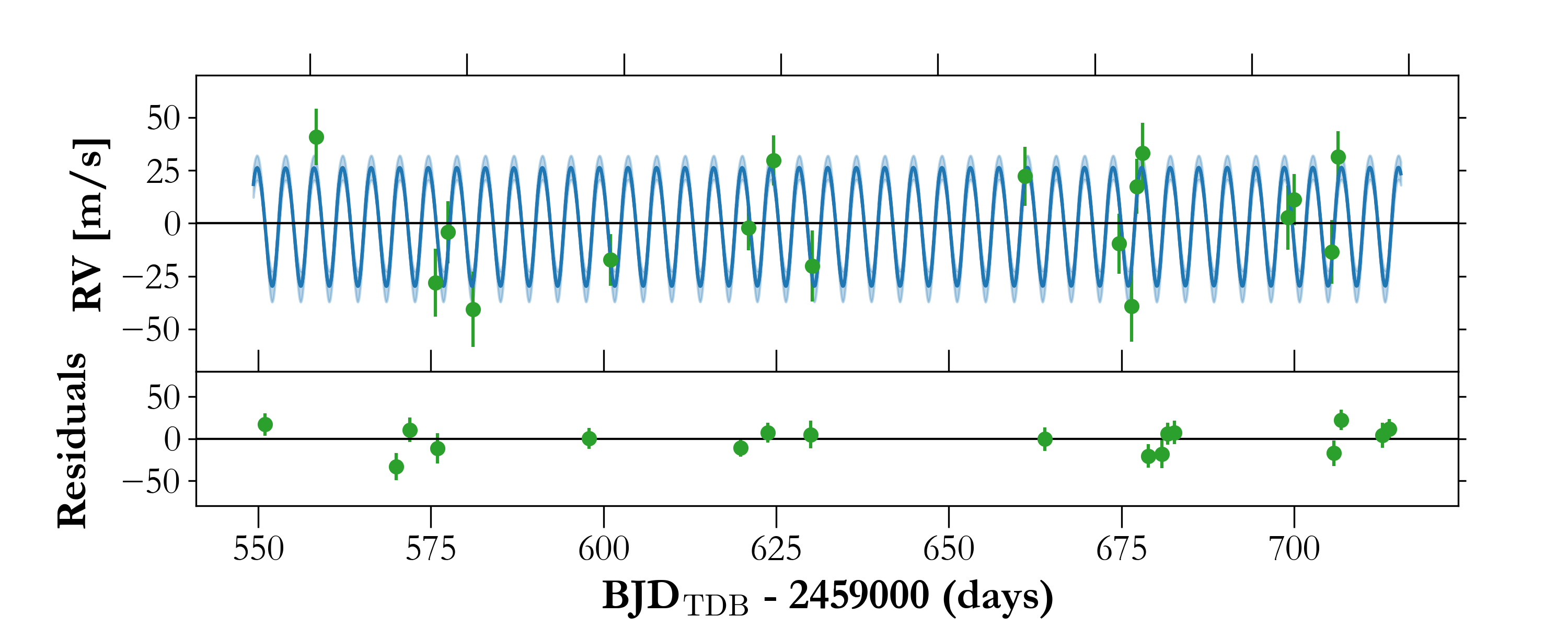}
    \caption{The full HPF RV time series with the best-fit model plotted in blue, with the 1-$\sigma$ quantile included as a lighter shade. }
    \label{fig:RVTimeSeries}
\end{figure*}

\begin{figure}[]
    \centering
    \includegraphics[width=0.5\textwidth]{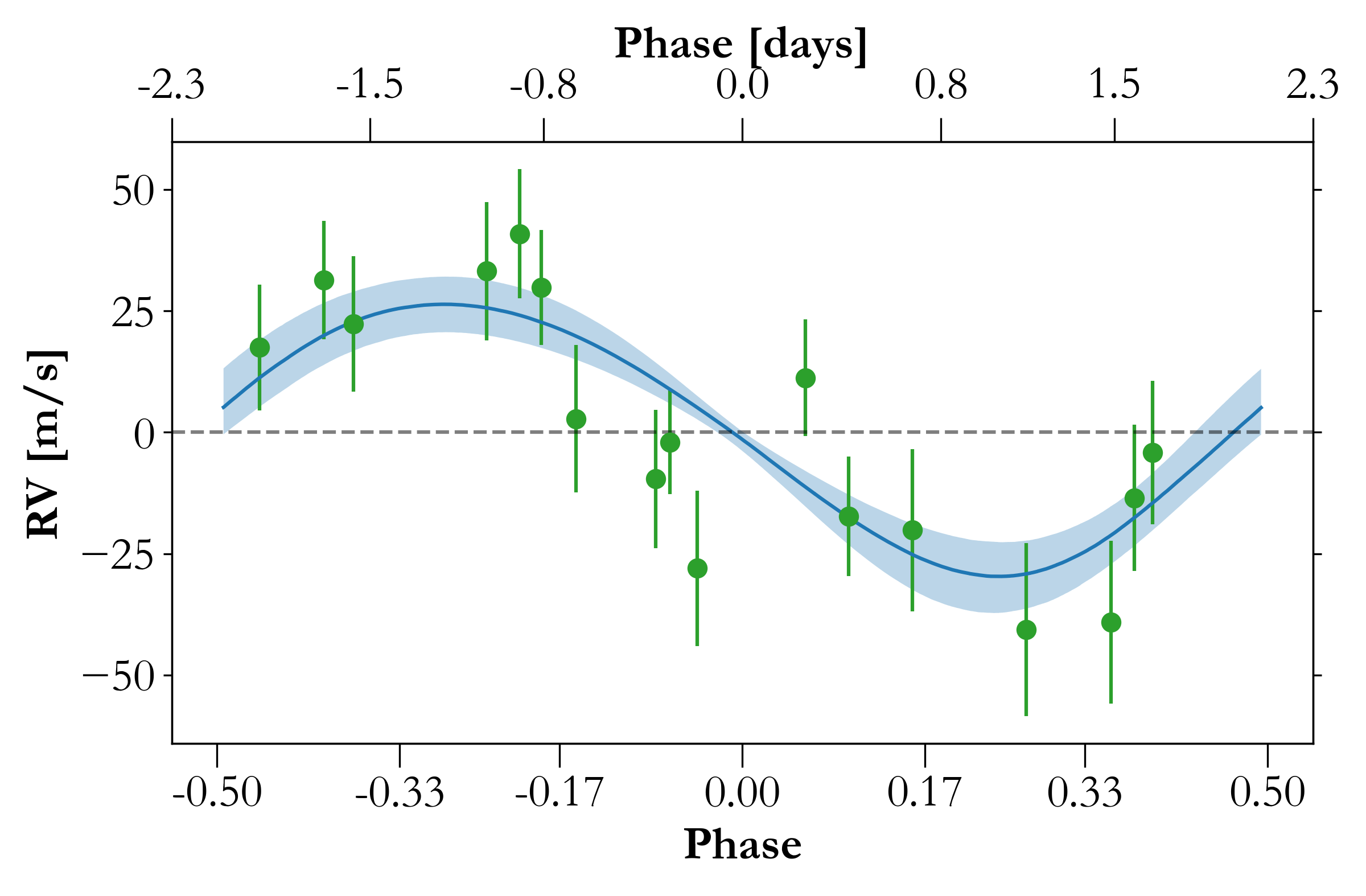}
    \caption{The HPF RVs phased to the best-fit period of TOI-3884b. Best-fit model and the 1-$\sigma$ quantile are plotted in blue. Mid-transit occurs at phase 0. }
    \label{fig:RVs}
\end{figure}

We removed all nights (8 total) which possessed unbinned S/N ratios less than 50\% of the expected S/N of 74 at 1.04 $\mu m$ calculated from the HPF Exposure Time Calculator\footnote{https://psuastro.github.io/HPF/Exposure-Times}. These S/N ratios ranged between 21 to 31. An inspection of these low S/N observations determined they were all obtained during less than optimal sky conditions (variable seeing $>$ 2 arcseconds, background i$^{\prime}$-band magnitude was brighter than 16.5, transparency was $<$ 75\% and/or bad weather or clouds were noted in the night logs). Every other observation possessed a S/N $>$ 43 and met our required observing conditions for transparency, seeing, and good weather conditions. We also removed the spectra from 2022 April 5 as these were observed during the transit spanning the large bump. As the planet is crossing an active region of the star, this may introduce potential contamination in the RV signal. This left 36 unbinned spectra taken over the course of 18 nights (Figures~\ref{fig:RVTimeSeries} and~\ref{fig:RVs}).

We applied a template-matching method \citep{anglada-escude_harps-terra_2012} using the SERVAL pipeline \citep{zechmeister_carmenes_2019} modified for HPF \citep{stefansson_mini-neptune_2020}. A master template was created by combining all spectra and masking tellurics and sky-emission lines. This template was then shifted to match each individual spectrum by minimizing the $\chi^{2}$ statistic before converting this shift into velocity space. We binned the two nightly individual RVs reported from SERVAL using a weighted-average based on their respective S/N ratios. The final binned RVs used for our analysis are listed in Table~\ref{tab:RVs} and are plotted in Figure~\ref{fig:RVs}.

\begin{deluxetable}{ccc}
\tablecaption{The $\sim$30 minute binned HPF RVs of TOI-3884. Low S/N points removed from the analysis are not included. \label{tab:RVs}}
\tablehead{\colhead{$\unit{BJD_{TDB}}$}  &  \colhead{RV}   & \colhead{$\sigma$}  \\
           \colhead{(d)}   &  \colhead{\ms{}} & \colhead{\ms{}}}
\startdata
2459550.99616          & 25                & 13                        \\
2459569.95174          & -44               & 16                        \\
2459571.94743          & -20               & 15                        \\
2459575.93822          & -56               & 18                        \\
2459597.88347          & -33               & 12                        \\
2459619.82335          & -18               & 11                        \\
2459623.80447          & 14                & 12                        \\
2459629.97493          & -36               & 17                        \\
2459663.88252          & 7                 & 14                        \\
2459678.84026          & -25               & 14                        \\
2459680.83590          & -55               & 17                        \\
2459681.65219          & 2                 & 13                        \\
2459682.64507          & 18                & 14                        \\
2459705.76008          & -13               & 15                        \\
2459706.76349          & -4                & 12                        \\
2459712.74903          & -29               & 15                        \\
2459713.74422          & 16                & 12                        \\
\enddata
\end{deluxetable}

\section{Analysis}\label{sec:analysis}
\subsection{Stellar Properties}\label{sec:stellar_properties}
We derived the spectroscopic stellar parameters: effective temperature (T$_{\rm{eff}}$), metallicity ([Fe/H]), and log g of TOI-3884 by applying the template matching methodology on the HPF spectra as outlined in \citet{stefansson_mini-neptune_2020}. Using the \texttt{HPF-SpecMatch} package \citep{stefansson_mini-neptune_2020}, we apply the spectral matching technique to the HPF Order 5 spectra (853 -- 864 nm) which has little to no telluric contamination. We list the spectroscopically derived stellar parameters for TOI-3884 from this analysis in Table~\ref{tab:stellarparam}. 

With the \texttt{isochrones} package \citep{isochrones}, we create an SED fit using the combination of the derived stellar spectroscopic values, the g, r, i, z, and y magnitudes reported from Pan-STARRS1 \citep[PS1;][]{chambers_pan-starrs1_2016,magnier_pan-starrs_2020}, the W1, W2, and W3 WISE-band magnitudes \citep{wright_wise_2010}, the J, H, and K magnitudes reported by 2MASS \citep{cutri_2mass_2033}, and the parallax from Gaia DR3 \citep{gaia.dr3.release}. We utilize Gaussian priors for all parameters except for a flat prior on the A$_{V}$ extinction and flat-log age prior up to 2 Gyr (see Section~\ref{sec:stellar_properties}). We utilize the relations in \citet{green_3d_2019}, calculated for the Gaia reported distance of 43 pc, to place an upper A$_{V}$ extinction limit of 0.1. We determine a stellar mass and radius of 0.298 $\pm$ 0.018 M$_{\odot}$ and 0.302 $\pm$ 0.012 R$_{\oplus}$ respectively (stellar density: 15.26 $\pm$ 2.04 g/cm$^{3}$) for TOI-3884. We verify these values by repeating the same fits using the \texttt{ExoFASTv2} package \citep{eastman_exofastv2_2019}, deriving masses and radii within 1$\sigma$. We verify this stellar density using the high-precision 2022 April 05 APO transit where we obtain a best-fit density of 15.43 $\pm$ 0.39 g/cm$^{3}$ (assuming a circular orbit). 

Using ESPRESSO, \citet{almenara.toi3884} suggests that TOI-3884 is a slowly rotating star with a slow \vsini~ of 1.1 km/s. They note this slow rotation suggests an inactive star -- in contrast with the large spot crossing event. We use our HPF spectra in an attempt to verify the slow rotator scenario by constraining the rotational broadening of TOI-3884 using two separate methods. First during the spectral-matching process, \texttt{HPF-SpecMatch} performs an optimization for the optimal rotational broadening \citep[see][for discussion]{stefansson_sub-neptune_sized_2020}. This results in a $v\sin i = 3.6\pm 0.9$ km/s. Second, we compare the widths of CCFs of TOI-3884 to the CCF widths of artificially broadened slowly rotating reference star of a similar spectral type. The \texttt{HPF-SpecMatch} analysis highlights Ross 128 as an excellent spectral match to TOI-3884b with a $T_{\mathrm{eff}}=3192 \pm 60$ K \citep{mann2015}, which matches well with the effective temperature of TOI-3884 of $T_{\mathrm{eff}}=3180 \pm 80$ K (Table \ref{tab:stellarparam}). Further, \citet{bonfils2018} demonstrate that Ross 128 is an inactive slowly rotating M dwarf with a long rotation period of $>100$ days, suggesting minimal rotational broadening. 

\begin{figure}
    \centering
    \includegraphics[width=0.5\textwidth]{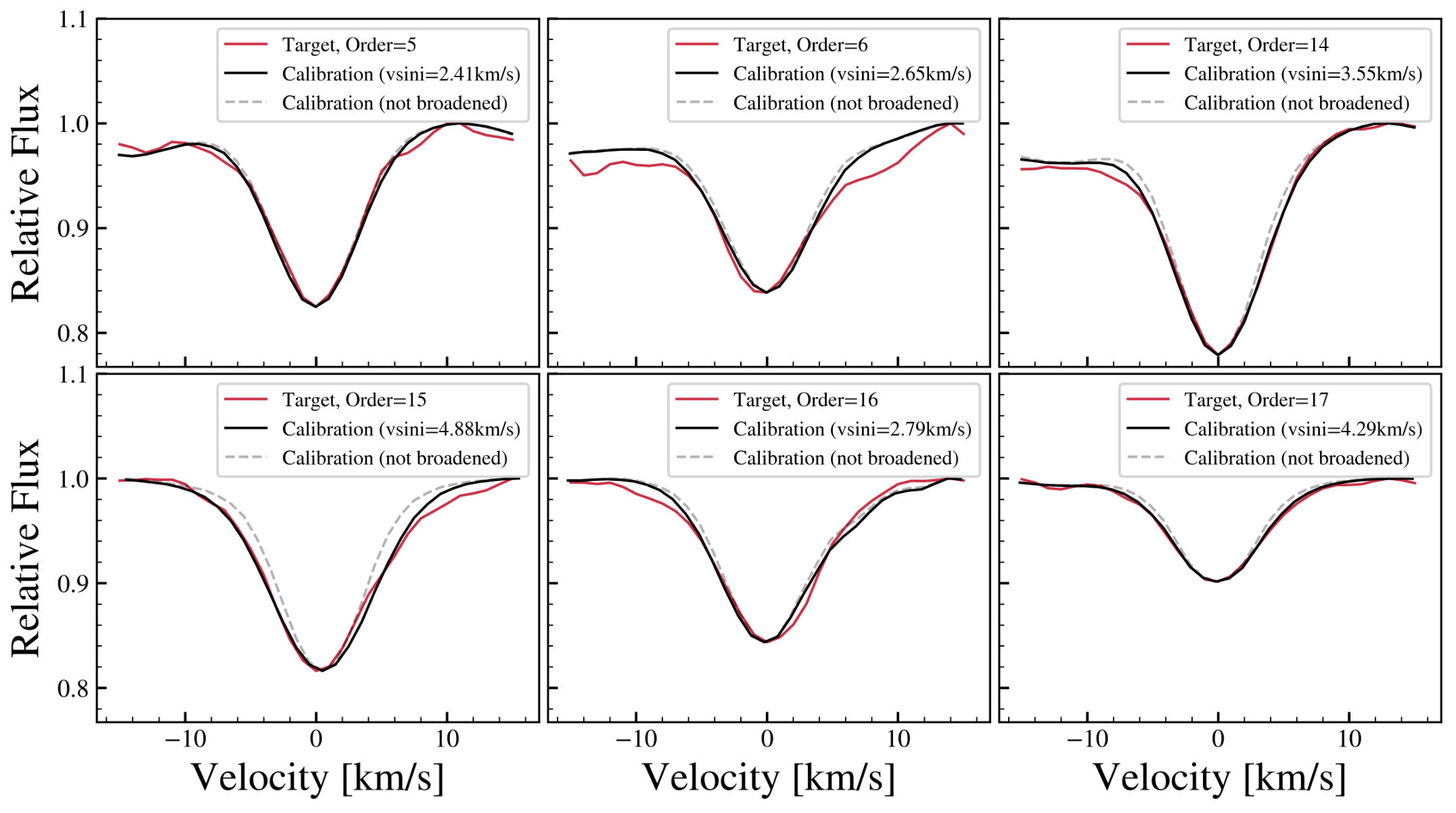}
    \caption{Comparing the width of the CCFs of TOI-3884 (red curves) in 6 different orders in HPF in 6 different panels to the CCFs of slowly rotating calibration star, Ross 128. The grey-dashed lines show the unbroadened calibration star, and the black lines show the calibration star artificially broadened to the best fit value. The TOI-3884 spectra show evidence for rotational broadening. }
    \label{fig:vsini}
\end{figure}

Figure \ref{fig:vsini} compares the CCFs of TOI-3884 to the CCFs of Ross 128 from 6 HPF orders clean of tellurics, suggesting that a $v \sin i > 3$km/s is warranted, and we derive a $v \sin i = 3.2 \pm 0.9$ km/s estimate from the average and the standard deviation values from the 6 HPF orders, respectively. We note that in trying to use other slowly rotating stars of similar spectral types results in similar $v\sin i$ values. We elect to formally adapt the $v\sin i$ value derived from \texttt{HPF-SpecMatch}, as through its $\chi^2$ minimization process of the full spectra it can better account for differences in normalization offsets that could lead to differences in the CCFs. We were unsuccessful to resolve the slower 1.1 km/s \vsini~ as originally published for this star.

The relatively rapid \vsini~from this work suggests TOI-3884 should be active and relatively young \citep[$<$ 1 Gyr;][]{newton_rotation_2016}. We support this conclusion with the LAMOST spectra which covers H$\alpha$. LAMOST reports an H$\alpha$ equivalent width (EW) of -3.86 $\pm$ 0.02 \AA~in emission. From Equation 1 in \citet{newton.halpha.relation}, an inactive star with TOI-3884's properties should have an H$\alpha$ EW of 0.18 \AA~ in absorption. \textit{TESS} observes three large flaring events in the two short cadence sectors, and the HPF spectra also show clear Ca IR triplet (Ca IRT) excess in emission.

We apply \texttt{pyHammer} \citep{roulston_pyhammer_2020} to the archival Large Sky Area Multi-Object Fiber Spectroscopic Telescope \citep[LAMOST;][]{lamost1,lamost2} spectra assuming the metallicity derived from the HPF spectra. Using a template-matching routine of empirical M dwarf spectra, we determine the best-fit spectral type to be either an M4 or M5 dwarf. We adopt M4 as the spectral type for this work.

Significant spot coverage can affect the measured photospheric temperature of the star and influence the SED derived stellar mass/radius. However, we detect no significant deviation from a single star SED fit to the observed magnitudes using \texttt{ExoFASTv2}. Moreover, we calculate a spot covering fraction of 15\% with a spot temperature of 2900 K will impact the actual stellar temperature by 50--100 K for TOI-3884. This is within the T$_\mathrm{eff}$ uncertainty reported by \texttt{HPF-SpecMatch}. Therefore, the derived stellar parameters in Table~\ref{tab:stellarparam} are minimally affected by the large spot (see Section~\ref{sec:starspot}).

\begin{deluxetable*}{lccc}
\tablecaption{Summary of stellar parameters for TOI-3884. \label{tab:stellarparam}}
\tablehead{\colhead{~~~Parameter}&  \colhead{Description}&
\colhead{Value}&
\colhead{Reference}}
\startdata
\multicolumn{4}{l}{\hspace{-0.2cm} Main identifiers:}  \\
~~~TOI & \tess{} Object of Interest & 3884 & \tess{} mission \\
~~~TIC & \tess{} Input Catalogue  & 86263325 & Stassun \\
~~~2MASS & \(\cdots\) &  J12061746+1230249 & 2MASS  \\
~~~Gaia DR3 & \(\cdots\) & 3919169687804622336 & Gaia DR3\\
\multicolumn{4}{l}{\hspace{-0.2cm} Equatorial Coordinates, Proper Motion and Spectral Type:} \\
~~~$\alpha_{\mathrm{J2016}}$ &  Right Ascension (RA) & 181.571808 & Gaia DR3\\
~~~$\delta_{\mathrm{J2016}}$ &  Declination (Dec) & +12.507030 & Gaia DR3\\
~~~$d$ &  Distance in pc  & 	
43.1 & Gaia DR3 \\
~~~\(A_{V,max}\) & Maximum visual extinction & 0.04 & Green\\
\multicolumn{4}{l}{\hspace{-0.2cm} Optical and near-infrared magnitudes:}  \\
~~~$B$ & Johnson B mag & 17.46 $\pm$ 0.23 & APASS\\
~~~$V$ & Johnson V mag & 15.74 $\pm$ 0.01 & APASS\\
~~~$g^{\prime}$ &  Sloan $g^{\prime}$ mag  & 16.62 $\pm$ 0.09 & Pan-STARRS1\\
~~~$r^{\prime}$ &  Sloan $r^{\prime}$ mag  & 15.17 $\pm$ 0.06 & Pan-STARRS1 \\
~~~$i^{\prime}$ &  Sloan $i^{\prime}$ mag  & 13.58 $\pm$ 0.06 & Pan-STARRS1 \\
~~~$J$ & $J$ mag & 11.13 $\pm$ 0.02 & 2MASS\\
~~~$H$ & $H$ mag & 10.55 $\pm$ 0.02 & 2MASS\\
~~~$K_s$ & $K_s$ mag & 10.24 $\pm$ 0.02 & 2MASS\\
~~~$W1$ &  WISE1 mag & 10.16 $\pm$ 0.02 & WISE\\
~~~$W2$ &  WISE2 mag & 9.99 $\pm$ 0.02 & WISE\\
~~~$W3$ &  WISE3 mag & 9.76 $\pm$ 0.05 & WISE\\
\multicolumn{4}{l}{\hspace{-0.2cm} \texttt{SpecMatch} Spectroscopic Parameters:}\\
~~~$T_{\mathrm{eff}}$ &  Effective temperature in \unit{K} & 3180 $\pm$ 88 & This work\\
~~~$\mathrm{[Fe/H]}$ &  Metallicity in dex & 0.04 $\pm$ 0.12 & This work\\
~~~$\log(g)$ & Surface gravity in cgs units & 4.97 $\pm$ 0.05 & This work\\
\multicolumn{4}{l}{\hspace{-0.2cm} Model-Dependent Stellar SED and Isochrone Fit Parameters:}\\
~~~$M_s$ &  Mass in $M_{\odot}$ & 0.298 $\pm$ 0.018 & This work\\
~~~$R_s$ &  Radius in $R_{\odot}$ & 0.302 $\pm$ 0.012 & This work\\
~~~$\rho_s$ &  Density in $\unit{g/cm^{3}}$ & 15.26 $\pm$ 2.04 & This work\\
\multicolumn{4}{l}{\hspace{-0.2cm} Other Stellar Parameters:}           \\
~~~$v \sin i_s$ &  Rotational velocity in \unit{km/s}  & 3.59 $\pm$ 0.92 & This work\\
~~~$P_{rot} (i_s = 90^\circ)$ &  Non-tilted maximum rotational period in \unit{days}  & 4.22 $\pm$ 1.09 & This work\\
~~~$\Delta$RV &  ``Absolute'' radial velocity in \unit{km/s} & 3.16 $\pm$ 2.89 & Gaia DR3\\
\enddata
\tablenotetext{}{References are: Stassun \citep{stassun_tess_2018}, 2MASS \citep{cutri_2mass_2033}, Gaia DR3 \citep{gaia.dr3.release}, Green \citep{green_3d_2019}, APASS \citep{henden_apass_2018}, WISE \citep{wright_wise_2010}, Pan-STARRS1 
\citep{chambers_pan-starrs1_2016,magnier_pan-starrs_2020}}

\end{deluxetable*}

\subsection{Joint Analysis of Transit and Radial Velocity Observations}\label{sec:planetsec}

We perform a joint-analysis of the transit and radial velocity (RV) observations to measure TOI-3884b's mass and radius. However, the lack of a pristine non-spot crossing transit light curve of TOI-3884b creates a challenge for transit analysis. We create an individualized starspot-mask based on visual inspection to each ground-based transit and to the three folded \textit{TESS} sectors. We then fit a transit model to the unmasked points. We calculate a $\chi^{2}_{r}$ of the residuals along with a by-eye examination. Points that still demonstrate bump-structure are masked and we repeat the procedure until we minimized the $\chi^{2}_{r}$. Masked duration and location vary slightly between data sets though all fell between 39 minutes prior to mid-transit (T$_0$) to 25 minutes post T$_0$; i.e, $\sim$67\% of the transit duration. Figure~\ref{fig:full_LCs} plots the ground-based and folded \textit{TESS} light curves with the masked points denoted in blue.

We perform a joint fit with \texttt{exoplanet} \citep{foreman-mackey_exoplanet_2021} using both the masked \textit{TESS} and ground-based transits and the HPF RVs. We fit for a single transit ephemeris, period, impact parameter, a/R$_{s}$, and transit depth using the combination of the three masked \textit{TESS} sectors, three TMMT transits, and three APO observations. We include a dilution term to the transit depth for each of the three \textit{TESS} sectors fixed on the ARCTIC SDSS i$^\prime$ transit. \texttt{exoplanet} uses the built-in \texttt{starry} \citep{luger_starry_2019} package to model the quadratic limb darkening parameters. As each instrument uses a different broadband filter, we fit for quadratic limb darkening terms specific to the various wavelength coverage (\textit{TESS} bandpass, Bessell I, SDSS i$^\prime$, and SDSS r$^\prime$). Last, we included a jitter term added in quadrature to the flux errors and a flux offset to each transit observation. Neither the \textit{TESS} out-of-transit baseline (PDCSAP) nor the ground-based observations required additional detrending. We plot our best-fit transit model in red for each transit in Figure~\ref{fig:full_LCs}.

We assume a Keplerian model for the RVs allowing eccentricity and the argument of periastron ($\omega$) to float as well as the RV semi-amplitude. Similar to the photometric transits, we include a jitter and RV offset terms as well as a general trend line. Including a GP had no effect on the RV results thus do not include an activity dependent GP for the RV orbit. We determine TOI-3884b is a super-Neptune with a mass of 32.59$^{+7.31}_{-7.38}$ M$_\oplus$ and radius of 6.43 $\pm$ 0.20 R$_\oplus$. Figure~\ref{fig:RVs} plots the best-fit RV model along with the 1$\sigma$ contours. We report the best-fit properties from this joint analysis as well as the final planetary properties for TOI-3884b in Table~\ref{tab:planetprop}. We note that the derived mass of TOI-3884b is based on the model's assumption that the planet is the main source of the RV variation. Periodograms of the Ca IRT, differential line widths, and chromatic index show no peaks with False Alarm Probabilities $<$ 10\% at the planet's period (nor any other period) indicating that the RV signal is not dominated by stellar activity.

\begin{deluxetable*}{llc}
\tablecaption{Derived Parameters for TOI-3884b.   \label{tab:planetprop}}
\tablehead{\colhead{~~~Parameter} &
\colhead{Units} &
\colhead{Value$^a$} 
}
\startdata
\sidehead{Orbital Parameters:}
~~~Orbital Period\dotfill & $P$ (days) \dotfill & 4.5445828 $\pm$ 0.0000098\\
~~~Eccentricity\dotfill & $e$ \dotfill & 0.06$^{+0.06}_{-0.04}$ \\
~~~Argument of Periastron\dotfill & $\omega$ (radians) \dotfill & -1.96$^{+4.28}_{-0.04}$ \\
~~~Semi-amplitude Velocity\dotfill & $K$ (\ms{})\dotfill &
28.03$^{+6.06}_{-6.23}$\\
~~~RV Trend\dotfill & $dv/dt$ (\ms{} yr$^{-1}$)   & 0.58$^{+4.78}_{-4.92}$   \\ 
~~~RV Jitter\dotfill & $\sigma_{\mathrm{HPF}}$ (\ms{})\dotfill & 7.86$^{+5.68}_{-5.11}$\\
\sidehead{Transit Parameters:}
~~~Transit Midpoint \dotfill & $T_C$ (BJD\textsubscript{TDB})\dotfill & 2459556.51669$\pm 0.00025$\\
~~~Scaled Radius\dotfill & $R_{p}/R_{s}$ \dotfill & 
0.197 $\pm$ 0.002\\
~~~Scaled Semi-major Axis\dotfill & $a/R_{s}$ \dotfill & 25.90$^{+0.96}_{-0.71}$\\
~~~Orbital Inclination\dotfill & $i$ (degrees)\dotfill & 89.81$^{+0.13}_{-0.18}$\\
~~~Impact Parameter\dotfill & $b$\dotfill &0.089$^{+0.082}_{-0.061}$\\
~~~Transit Duration\dotfill & $T_{14}$ (days)\dotfill & 0.0666$^{+0.0019}_{-0.0024}$\\
~~~Dilution$^{b}$ \dotfill & $D_{\mathrm{TESS~S22}}$ \dotfill & 0.98 $\pm$ 0.12\\
~~~ & $D_{\mathrm{TESS~S46}}$ \dotfill & 0.86 $\pm$ 0.03\\ 
~~~ & $D_{\mathrm{TESS~S49}}$ \dotfill & 0.84 $\pm$ 0.03\\ 
\sidehead{Planetary Parameters:}
~~~Mass\dotfill & $M_{p}$ (M$_\oplus$)\dotfill &  $32.59^{+7.31}_{-7.38}$\\
~~~Radius\dotfill & $R_{p}$  (R$_\oplus$) \dotfill& 6.43 $\pm$ 0.20\\
~~~Density\dotfill & $\rho_{p}$ (\gcmcubed{})\dotfill & 0.67$^{+0.18}_{-0.16}$\\
~~~Semi-major Axis\dotfill & $a$ (AU) \dotfill & $0.0361\pm0.0008$\\  
~~~Planetary Insolation & $S$ (S$_\oplus$)\dotfill &  6.29 $\pm$ 0.84\\
~~~Equilibrium Temperature$^c$ \dotfill & $T_{\mathrm{eq}}$ (K)\dotfill & 441 $\pm$ 15\\
\enddata
\tablenotetext{a}{The reported values refer to the 16-50-84\% percentile of the posteriors.}
\tablenotetext{b}{Dilution due to presence of background stars in \tess{} aperture, not accounted for.}
\tablenotetext{c}{We assume the planet to be a black body with zero albedo and perfect energy redistribution to estimate the equilibrium temperature. }
\normalsize
\end{deluxetable*}

\section{Starspot Analysis}\label{sec:starspot}

We now focus on analyzing the ubiquitous spot feature that is present in all the transit light curves shown in Figure~\ref{fig:full_LCs}. In-transit flux increases like this one are commonly observed in planetary transits, when the planet passes in front of a localized region of reduced flux on the surface of the star (i.e. a starspot) \citep{Rabus2009,sanchis_ojeda_hatp11_2011,morris2017,schutte_starspot_2022}. Precise knowledge of the planet's orbital properties in combination with the stellar rotation and tilt can provide specific positional information about the spots in the path of the planet as shown in \cite{morris2017,schutte_starspot_2022}. Conversely, observations of multiple in-transit spot occultations combined with inferences about the spot properties can provide information about the obliquity of the planet, as found in \cite{sanchis_ojeda_hatp11_2011}.  It is important to note that while we refer to the spots as "spots" what we are most likely modeling are entire spot complexes. 

TOI-3884 shows a prominent star spot crossing feature in every high cadence transit between December 2021 and June 2022, and a single point flux increase in-transit in the long-cadence \textit{TESS} light curve from Spring 2020.  While spot occultations are often detected in multiple, different transits of the same star, TOI-3884 is unique in that the feature persists at the same orbital phase in the first half of the transit for at least two years. The similarity of the amplitude, duration, and shape of the features combined with its persistence suggests that we are observing the same long-lived spot in all the light curves.  

There is a very limited parameter space of stellar rotation and stellar inclination that would result in the same spot being detected at the same orbital phase over the 6-month duration of the observations, given the well-defined orbital period of the planet. If the star is spinning upright (i.e. the stellar inclination is 0$^{\circ}$), the persistent spot could only occur if the star was rotating so slowly that the spot appeared fixed in place (which is incredibly unlikely over two years of monitoring), or if the orbital period of the planet was an exact integer multiple of the star's rotation period so that each time the planet transited, a spot feature was back in the same location relative to the observer. The only other scenario which would result in the fixed phase of the star spot feature is one in which the star is tilted away from the line of sight and the large persistent spot is fixed at or near the pole so that it doesn't move relative to the observer even as the star rotates. Based on the measured \vsini, we rule-out the slow rotating scenario. However, we explore the other two possibilities: i) a non-tilted star with synchronous rotation and ii) a tilted star system with a non-zero obliquity. For both possible scenarios, we keep the transit parameters fixed to the ones found in Section \ref{sec:planetsec}.

\subsection{Starspot Model of a Non-Tilted Star with Synchronous Rotation}

We apply the program STarSPot \citep[\texttt{STSP};][]{morris2017, schutte_starspot_2022} to the high-precision 2022 April 05 APO observation with the procedure outlined in \cite{schutte_starspot_2022}. \texttt{STSP} is specifically designed to model the light curves of transiting systems in which star spots and/or faculae create localized surface brightness variations on the host star.  Using an affine-invariant Markov Chain Monte Carlo (MCMC) \citep{morris2017} optimizer, a single run of \texttt{STSP} samples different radii (R$_{\rm spot}/$R$_{s}$), latitudes, and longitudes ($\theta$ and $\phi$ respectively) for every spot, but applies a fixed spot contrast (defined by its temperature and the filter of the observed transit) in order to break the known degeneracies between these properties.  

For TOI-3884, we assume a photospheric temperature of 3200~K derived from the HPF spectroscopic parameters and perform \texttt{STSP} runs with the following set of spot temperatures: 2600, 2700, 2800, 2900, 3000, and 3100 K. We calculate the contrasts defined by these temperatures by first interpolating \texttt{PHOENIX} synthetic \citep{phoenix} spectra at both the stellar and spot temperatures.  We then integrate over the specific filter response curve of the observed light curve and sum the integrated flux in that filter. Finally, we calculate the contrast for each spot temperature by dividing the integrated spot flux by the integrated photospheric flux (Schutte et al. In Prep). 

We first consider the star-planet orientation in which the spin-axis of the star is in the plane of the sky and aligned with the orbital axis of the planet.  In this case, the measured \vsini~provides a rotation period of $P_{\rm rot} = 4.22 \pm 1.09 $~d, which is consistent with the orbital period ($P_{\rm orb} = 4.54$ d).  
Adopting a rotation period for the star of $P_{\rm rot} = 4.54$ d that is synchronous with the orbital period and applying the \texttt{STSP} program to the APO i$^\prime$~band transit, we quickly find that a single circular starspot is insufficient to describe the structure of the feature, but a three spot model, with one large spot surrounded by two smaller spots produces the lowest $\chi^2$, regardless of spot contrast. Even after forcing the model to have two medium-sized spots in the place of the large central spot, the optimization preferred one large pole-spot combined with the two smaller nearby spots.

\begin{figure}[h]
    \centering
    \begin{minipage}{0.48\textwidth}
        \centering
        \includegraphics[width=\textwidth]{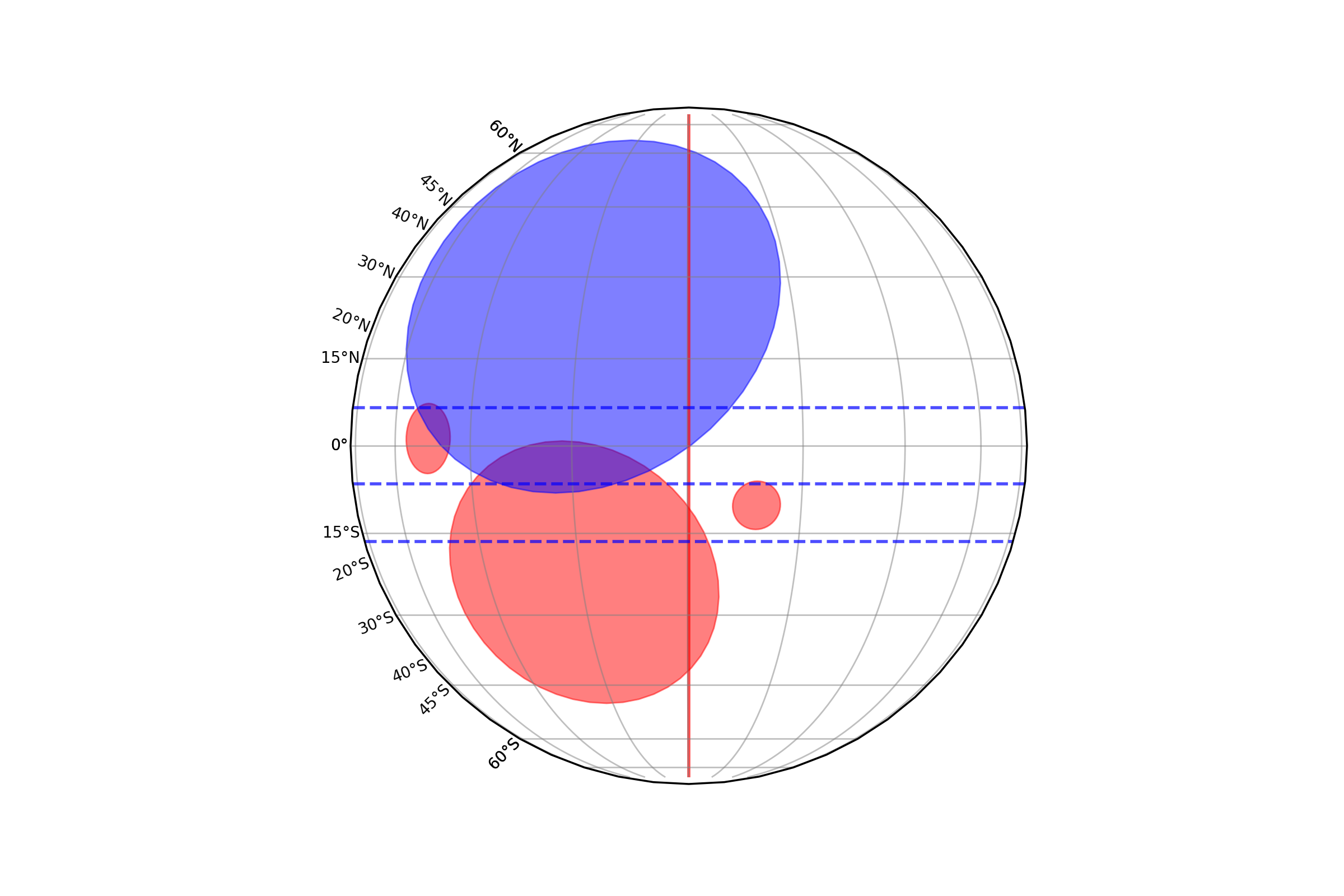} 
    \end{minipage}%
    \begin{minipage}{0.48\textwidth}
        \centering
        \includegraphics[width=\textwidth]{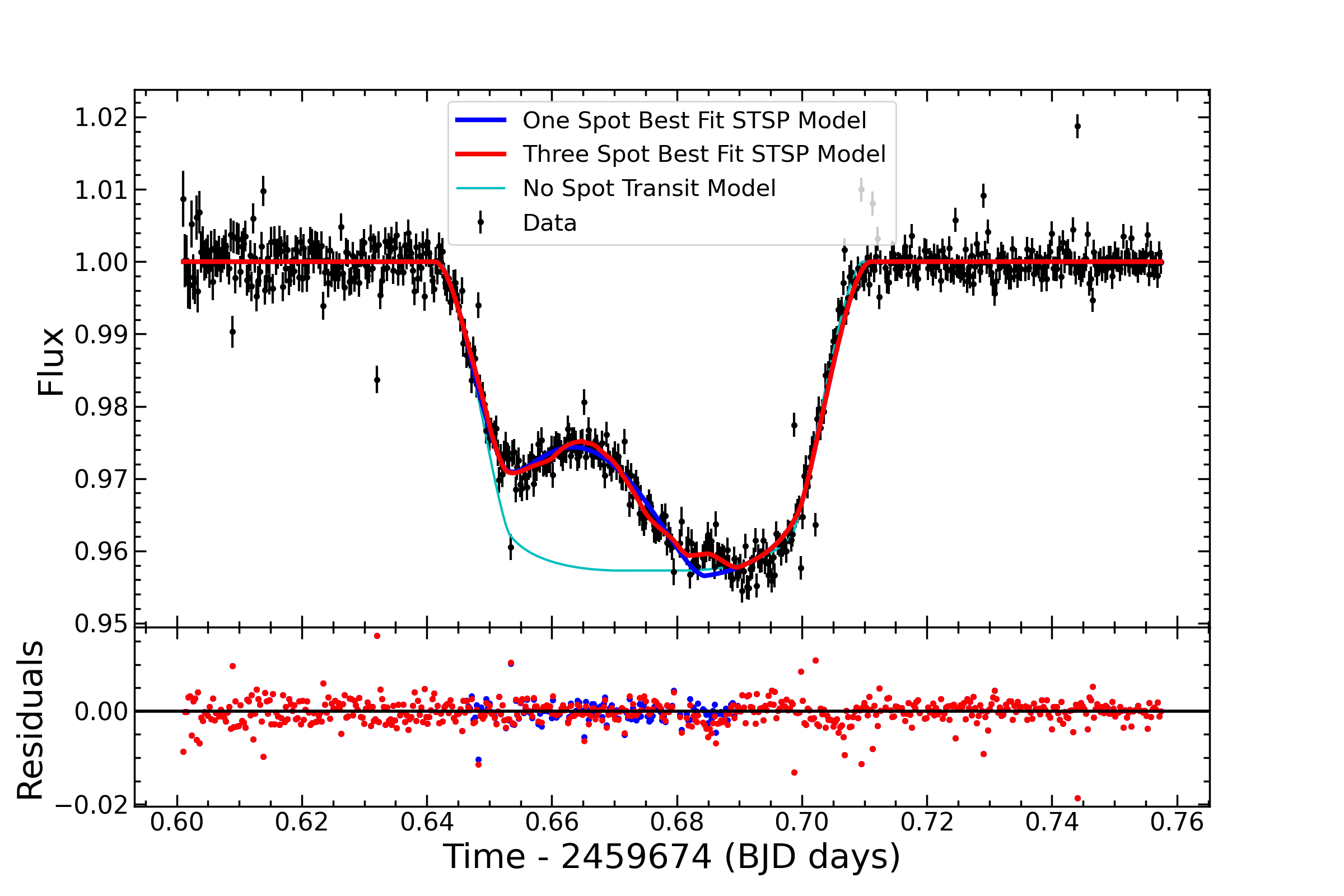} 
        \caption{\textbf{Top:} Projected starspots on TOI-3884's stellar surface for an aligned system using the APO \(i^\prime\) filter transit observed on 2022 April 05 assuming a spot temperature of 2900 K and a photospheric temperature of 3200 K (spot contrast of 0.5) for the three spot model (red) and one spot model (blue).  The planet's crossing path is defined by the blue dotted lines with the central dotted line corresponding to the equator of planet and the outer dotted lines denoting the full extent of the planet, and the central latitude of the transit is marked with a red vertical line.  The large red spot in the middle has a relative radius R$_{\rm spot}$  = 0.44 R$_s$ with the two smaller red spots having radii R$_{\rm spot}$  = 0.10 and 0.07 R$_s$ respectively. The large blue spot has a relative radius R$_{\rm spot}$  = 0.63 R$_s$ and is mostly out of the transit chord. The fractional spotted area for the three spot model in the transit chord for these stellar surface features (assuming there are no spots anywhere else on the star) is 11\%. \textbf{Middle:} Best fit three spot model for aligned system shown in red line compared to the best fit one spot model in blue line and the no star spot transit model in cyan with the APO 20s \(i^\prime\) band data as black points with error bars. \textbf{Bottom:} Residuals from the three spot best-fit starspot model (red) and one spot best-fit model (blue).}\label{fig:sphere_align}
    \end{minipage}
\end{figure}

The best-fit three spot model is shown in Figure~\ref{fig:sphere_align}, consisting of one large central spot (R$_{\rm spot}/$R$_s$ = 0.44) surrounded by two smaller spots (R$_{\rm spot}/$R$_s$ = 0.10 and 0.07 respectively). This model has a reduced $\chi^2$ of 2.14 and corresponds to a spot temperature of 2900 K (contrast$~=~0.5$). In comparison, the best fit one and two-spot models have reduced $\chi^2$ values of 2.25 and 2.16 respectively. If we compare the AIC$_c$ \citep{sugiura1978} values between the one, two, and three spot models (-9.24, -6.12, and -2.86 respectively), the AIC$_c$ favors the one-spot model as it has the least amount of fitting parameters. It is important to note that both the one and two-spot models do not fit the data points as well by eye (the one spot model is shown in Figure \ref{fig:sphere_align}). Therefore, even though the AIC$_c$ favors the simplest one spot model, the $\chi^{2}$ statistic prefers the three-spot model. Additionally, the spot feature is asymmetric, which is hard to fit with fixed circles as is required by \texttt{STSP}, but if we instead have two spots close together, this can replicate an asymmetric feature, which further adds to the three spot model being the best-fit model. Spot temperatures of 2700 and 2800 K produce equally good solutions (reduced $\chi^2$ values of 2.19 and 2.17 respectively which fall within $\sigma_{\chi^2} = 34$) with similar spot configurations. While the reduced $\chi^2$ values are larger than one, the data are ground-based data with likely underestimated error bars. Hotter spot temperatures (3000 and 3100 K) cannot reproduce the amplitude of the feature and are therefore not possible solutions regardless of the spot configuration. The coolest spot temperature of 2600 K produces a similar spot configuration, but the reduced $\chi^2$ falls just outside of the above variance. Given the discrete 100 K sampling of our models, we find the temperature of the large spot to be between 2700 to 2900 K, with a preference for the hotter 2900 K spot. From these fits we constrain the radius of the large spot to be R$_{\rm spot}/$R$_s$ = 0.44 $\pm$ 0.08. 

\begin{figure*}
\centering
 \includegraphics[width=\textwidth]{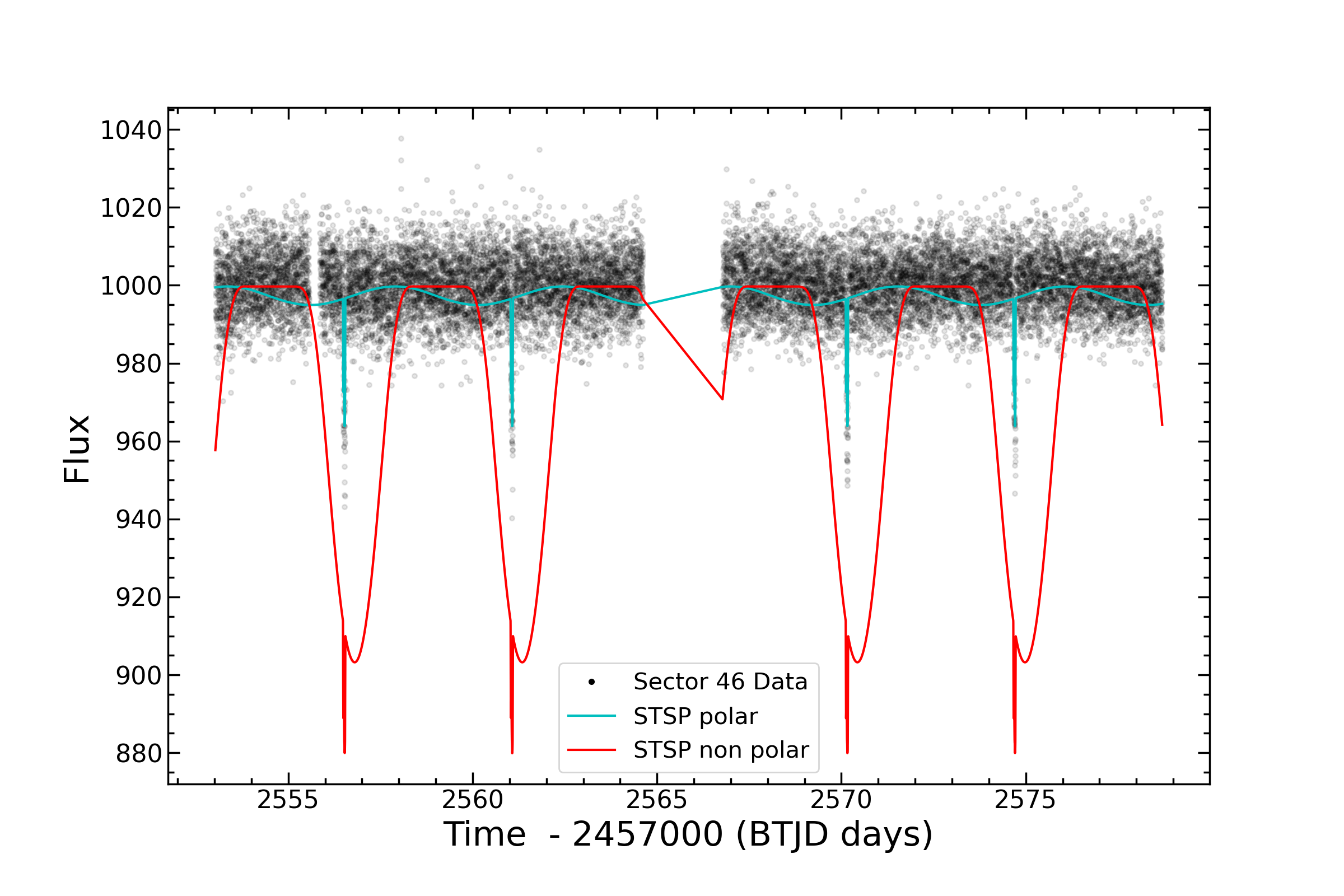}
 \label{fig:comp_s46}

\caption{TESS Sector 46 short cadence (2 minute) data shown in black points with aligned starspot model (red line) and polar starspot model (cyan line). The same pattern can be seen in TESS Sector 49 short cadence data though it is not reproduced here.}
\label{fig:comp_tess}
\end{figure*}

This scenario produces a very large (radius of $\sim 44\%$ the star's radius) star spot which will produce significant photometric out of transit variability of $>$ 1\% if that is the only large feature on the star. Interestingly, we do not detect any clear photometric modulations in the two short-cadence TESS sectors nor in the publicly available ground-based monitoring with Zwicky Transient Facility \citep[ZTF;][]{Masci2019}, All-Sky Automated Survey for Supernovae \citep[ASAS-SN;][]{asassn.citation}, and the Asteroid Terrestrial-impact Last Alert System \citep[ATLAS;][]{atlas.citation}. Figure~\ref{fig:comp_tess} shows \textit{TESS} Sector 46 data as an example with Sector 49 showing the same pattern. To explore this scenario fully, we model the full light curve for both \textit{TESS} Sector 46 and 49 with the in-transit star spots fixed but with an additional three spots allowed to vary. We found that in order to decrease the out of transit variability enough to be less than the noise level ($< 0.5\%$) of the \textit{TESS} light curves, there must be additional spots such that as the star rotates there is always a near equal fraction of spotted area (20\%) rotating out of view as is rotating into view. Thus, it is  possible the photometric spot modulation of TOI-3884 is simply hidden within the noise assuming the spots are configured such that they are uniformly spread across the surface of the star and cover a large fraction of the star. 

This leads to the concern that the RV-observed 4.56 d signal is partially due to the stellar rotation and not the planet. However, none of the HPF activity indicators show any periodic signal which would be characteristic for large spots \citep{robertson_persistent_2020}.

While observations cannot formally exclude this scenario, the requirements are contrived: TOI-3884 must have a rotation exactly equal to its planet, possess a spotted surface such that the photometric variability is $<$ 0.5\% over the two \textit{TESS} sectors, maintain the same starspot with very little evolution across in the transit chord while keeping the rest of the transit chord nearly spot-free. We therefore disfavor this hypothesis.

\subsection{Model of Tilted Star System with Non-Zero Obliquity}

Tilting the star such that the spot does not rotate in and out of view would minimize the out-of-transit variability \citep{jackson_spotmodulation_2012}. In this scenario, TOI-3884's large spot must be located on or near the pole of the star with the star's spin axis inclined ($i_s$) such that the pole of the star is pointed towards the observer. In order for the spot feature to occur at the same phase in the first half of the transit, the spin axis of the star and the planet's orbital axis must be misaligned (i.e. $\lambda$ $\neq 0$) . 
Because \texttt{STSP} assumes the planet's position and the star's tilt are well known, it is not designed to derive the optimal stellar inclination and $\lambda$. However, the fixed phase of the spot feature allows us to constrain the tilt of the star's spin axis and $\lambda$. For example, if the pole of the star is pointed exactly towards the observer (stellar inclination of $0^{\circ}$), the bump would occur in exactly the middle of the transit regardless of $\lambda$. Conversely, a tilt that is too close to the plane of the sky (i.e. inclination $> 60^{\circ}$) produce spot crossings during ingress of the transit. Thus, we first performed a comprehensive search of every stellar inclination value from $60^{\circ}$ down to exactly pole on in increments of $5^{\circ}$. For our search, we fixed the rotation period to be exactly equal to the orbital period, and we assumed one spot with a radius of 30\% the radius of the star was directly on the Southern pole of the star. From our search, we found that the only possible if $i_s$ was $25^{\circ} \pm 5^{\circ}$. After determining the best stellar inclination for the star, we then performed a series of simulations which varied $\lambda$ from $0^{\circ}$ to $180^{\circ}$ in increments of $10^{\circ}$. From our search, we determined that $\lambda = 75^{\circ} \pm 10^{\circ}$ provided the best fit to the APO SDSS i$^\prime$ light curve. It is important to note the provided uncertainties were derived from an exploration of the possible stellar inclinations that fit the data, and then, the uncertainties for $\lambda$ assume the stellar inclination is constant. Since these parameters are actually entwined, a more formal determination of the error bars is left to future work.

\begin{figure}[h]
    \centering
    \begin{minipage}{0.48\textwidth}
        \centering
        \includegraphics[width=\textwidth]{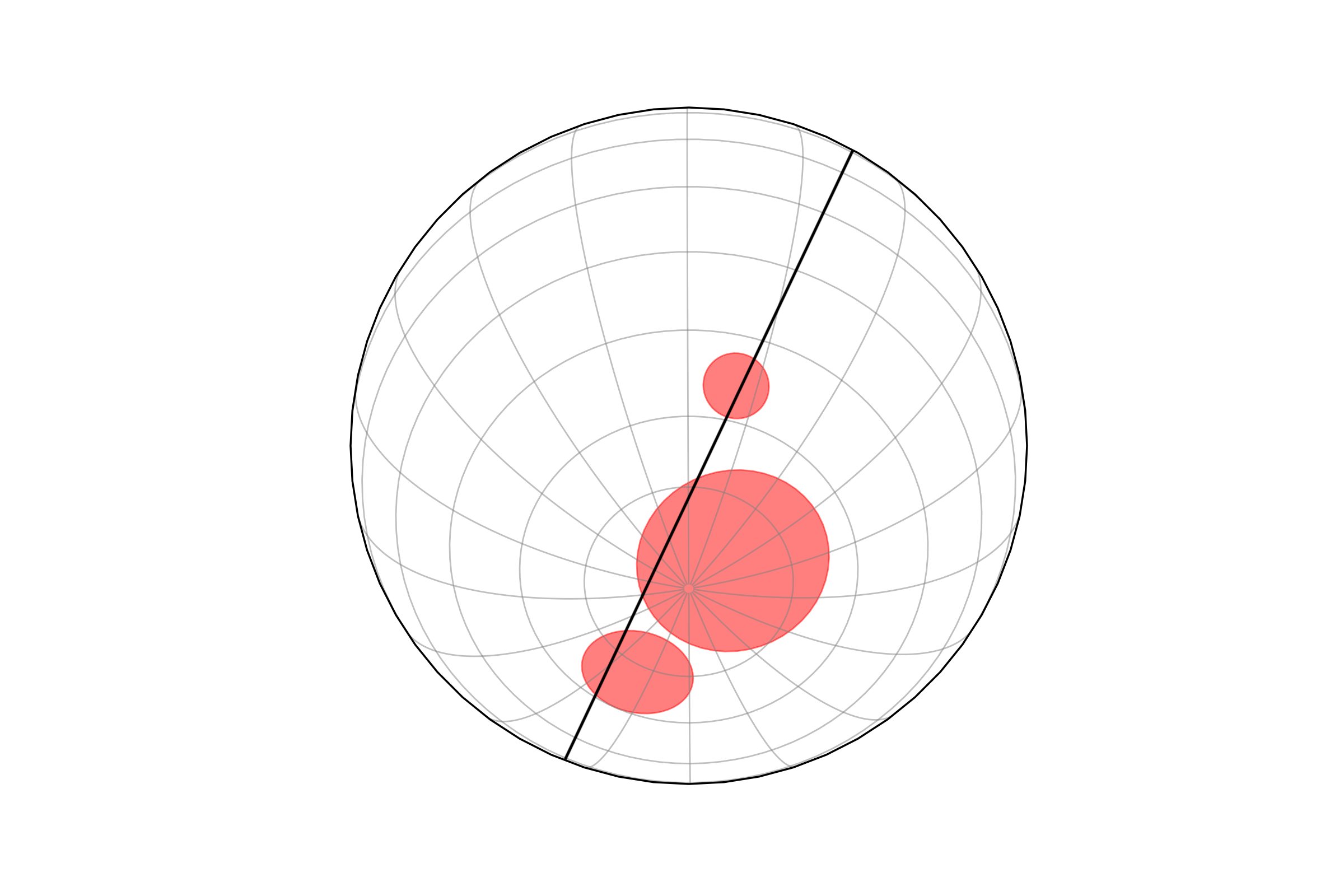} 
    \end{minipage}%
    \begin{minipage}{0.48\textwidth}
        \centering
        \includegraphics[width=\textwidth]{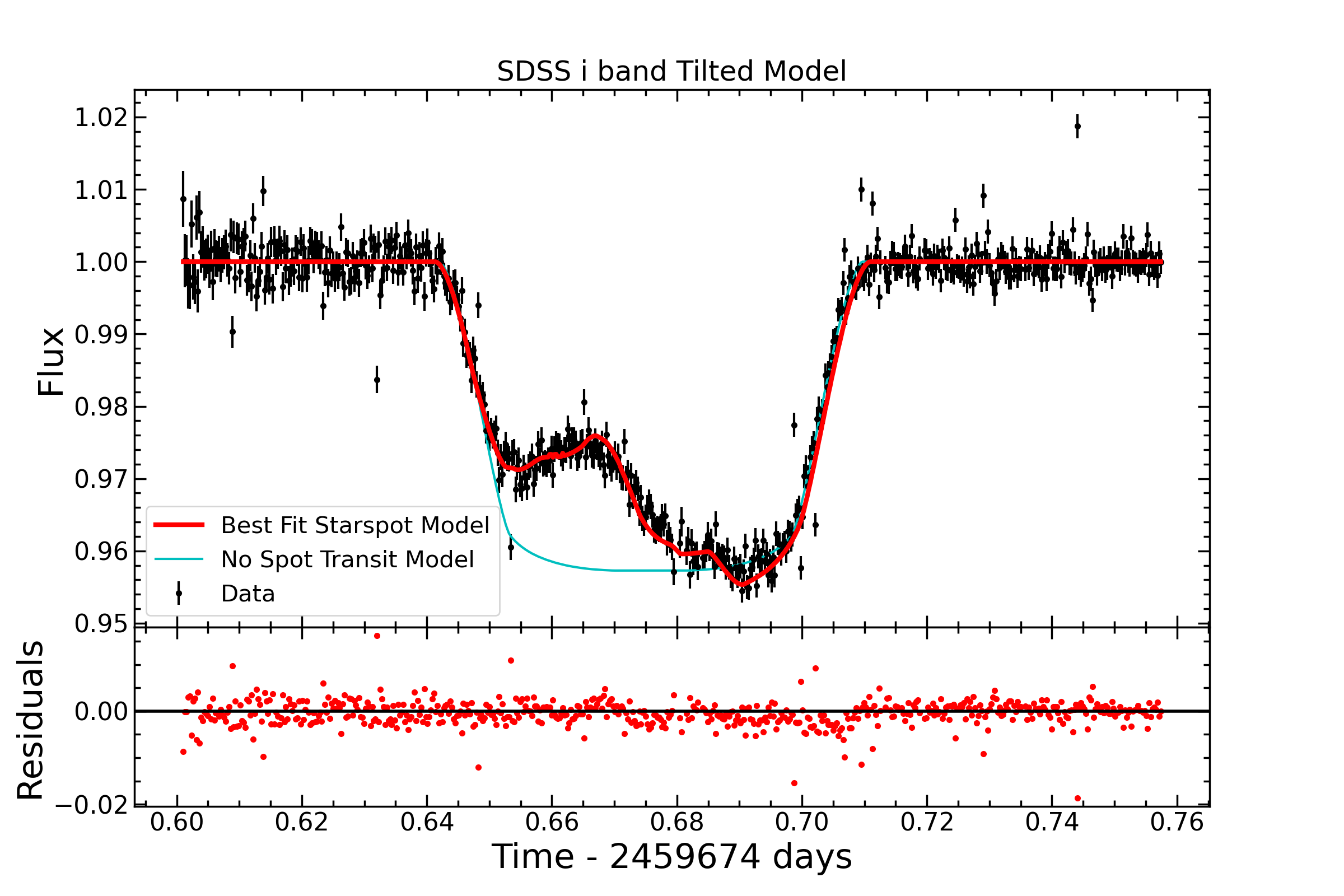} 
        \caption{\textbf{Top:} Projected star spots on TOI-3884's stellar surface for a polar star spot with a stellar spin axis tilt of $-65^{\circ}$ and $\lambda = 75^{\circ}$ using the APO SDSS \(i^\prime\) filter transit observed on 2022 April 05 assuming a spot temperature of 2900 K and a photospheric temperature of 3200 K (spot contrast of 0.5). The large spot in the middle has a relative radius R$_{\rm spot}/$R$_s$  = 0.29 with the two smaller spots having radii R$_{\rm spot}/$R$_s$  = 0.16 and 0.09 respectively. The fractional spotted area in the transit chord for these stellar surface features assuming there are no spots anywhere else on the star is 3\%. The black line shows the path of the equator of the planet as it crosses the star. \textbf{Bottom:} Best fit starspot model for the oblique (not aligned) system shown in red line compared to the no starspot transit model in cyan with the APO 20s \(i^\prime\) band data as black points with error bars.}\label{fig:sphere_polar}
    \end{minipage}

\end{figure}

Once we determined the stellar spin axis and $\lambda$ for the misaligned scenario, we modeled the star spots in the same way as before with \texttt{STSP} where the radii and locations of the spots are allowed to vary. For this scenario, we assume the same number of spots and spot temperature as found for the best-fit aligned model (three spots with temperatures of 2900 K) as it is likely the spot temperature and number of spots is the same no matter the tilt of the star. We found the best fit stellar surface features for this scenario to be one large spot (R$_{\rm spot}$/R$_s$ = 0.29) that is slightly off-center to the pole with two smaller spots on either side (R$_{\rm spot}/$R$_{s}$ = 0.16 and 0.09 respectively). This spot configuration is shown in Figure~\ref{fig:sphere_polar} and has a final reduced $\chi^2$ of 2.6. We also fit this scenario with one large spot instead of three spots, though the reduced $\chi^2$ of this model was 3.96. Similar to the results with the non-tilted star, the AIC$_c$ value for the one-spot model is lower than the three spot model (-10.38 and -3.04 respectively) due to the difference in parameters. The one-spot model does not fit the data well by eye, and again the $\chi^{2}$ statistic favors the three spot model. Therefore, we opt for the three spot model.

Finally, we calculate the out of transit variability for the \textit{TESS} short cadence data and find it produces variability well below the \textit{TESS} noise level with no additional large spots needed (see Figure~\ref{fig:comp_tess}). Table~\ref{tab:spot} reports the best-fit values from this analysis.

\begin{table}[]
\caption{Parameters Derived from Pole-spot Model }
\begin{tabular}{lc}
\hline
Spot Temperature Range                    & [2700, 2900] K  \\
Spot Radii (R$_\textrm{spot}$/R$_s$) & 0.29, 0.16, 0.09    \\
Stellar Inclination ($i_{s}$)$^a$         & 25 $\pm$ 5$^\circ$  \\
Sky-Projected Obliquity ($\lambda$)$^a$  & 75 $\pm$ 10$^\circ$ \\ \hline
\end{tabular} 
\footnotesize{\newline $^a$ Uncertainties derived independently of one another.} \label{tab:spot}
\end{table}

We use \texttt{SOAPv2} \citep{dumusque.soap.2014} to test the impact of a pole-spot on the RVs. A pole-spot under this scenario will inject a $\sim$10 m/s signal, which is 30\% of the RV semi-amplitude. However, \texttt{SOAPv2} assumes an optical bandpass as it was designed specifically for the HARPS wavelength range (380 -- 700 nm). Stellar activity decreases at longer wavelengths where the contrast between the spot and photosphere temperatures decreases \citep{reiners.nIR.RV.2010}. As HPF operates at near-infrared wavelengths, we expect the overall impact of the spot to be suppressed by $\sim$ 2$\times$ \citep{reiners.nIR.RV.2010,robertson_persistent_2020}. Thus, the pole-spot's impact with this configuration is $<$ 5 m/s -- within the 1$\sigma$ semi-amplitude uncertainty of 28.0 $\pm$ 6.3 m/s. 

\subsection{Evidence for Spot-Complex Evolution} \label{subsec:evolution}

We extend our spot model derived from the SDSS i$^\prime$ transit to the APO SDSS r$^\prime$ and \textit{TESS} short cadence observations. We calculate new contrast values for the \textit{TESS} and SDSS r$^\prime$ band filters for a spot temperature of 2900 K. Using the exact spot-complex configuration for the misaligned system, we model the r$^\prime$ band and TESS short cadence transits using \texttt{STSP}. The results showed that the same spot configuration could not fit either the \textit{TESS} or APO r$^\prime$ band transits. 

A close inspection of individual \textit{TESS} transits suggest \textit{slight} changes in spot amplitude, duration, and location (though it always starts during ingress). However, the lack of precision within the individual \textit{TESS} transits makes it near-impossible to map spot evolution of subsequent transits. 

\begin{figure}[h]
    \centering
    \begin{minipage}{0.48\textwidth}
        \centering
        \includegraphics[width=\textwidth]{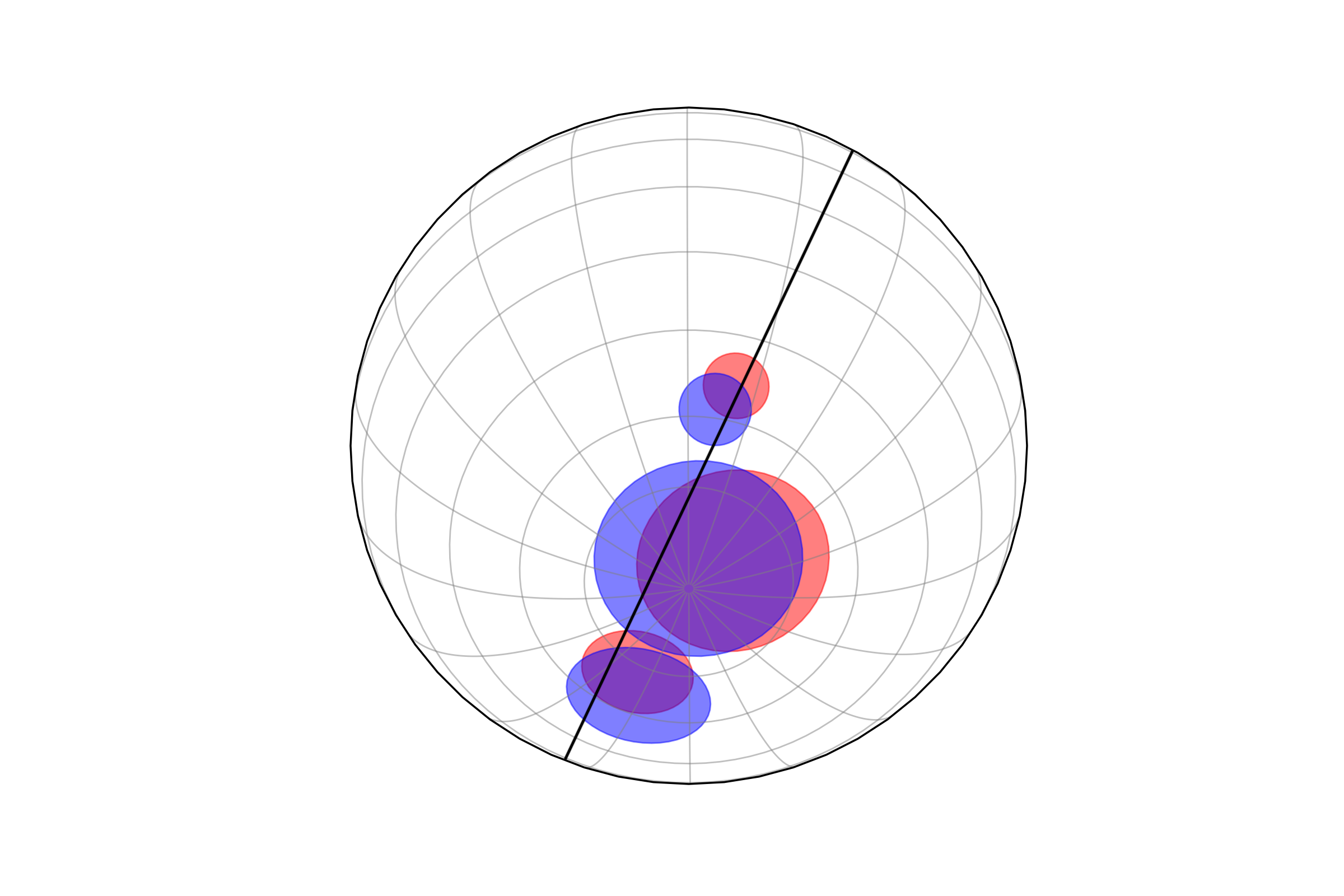} 
    \end{minipage}%
    \begin{minipage}{0.48\textwidth}
        \centering
        \includegraphics[width=\textwidth]{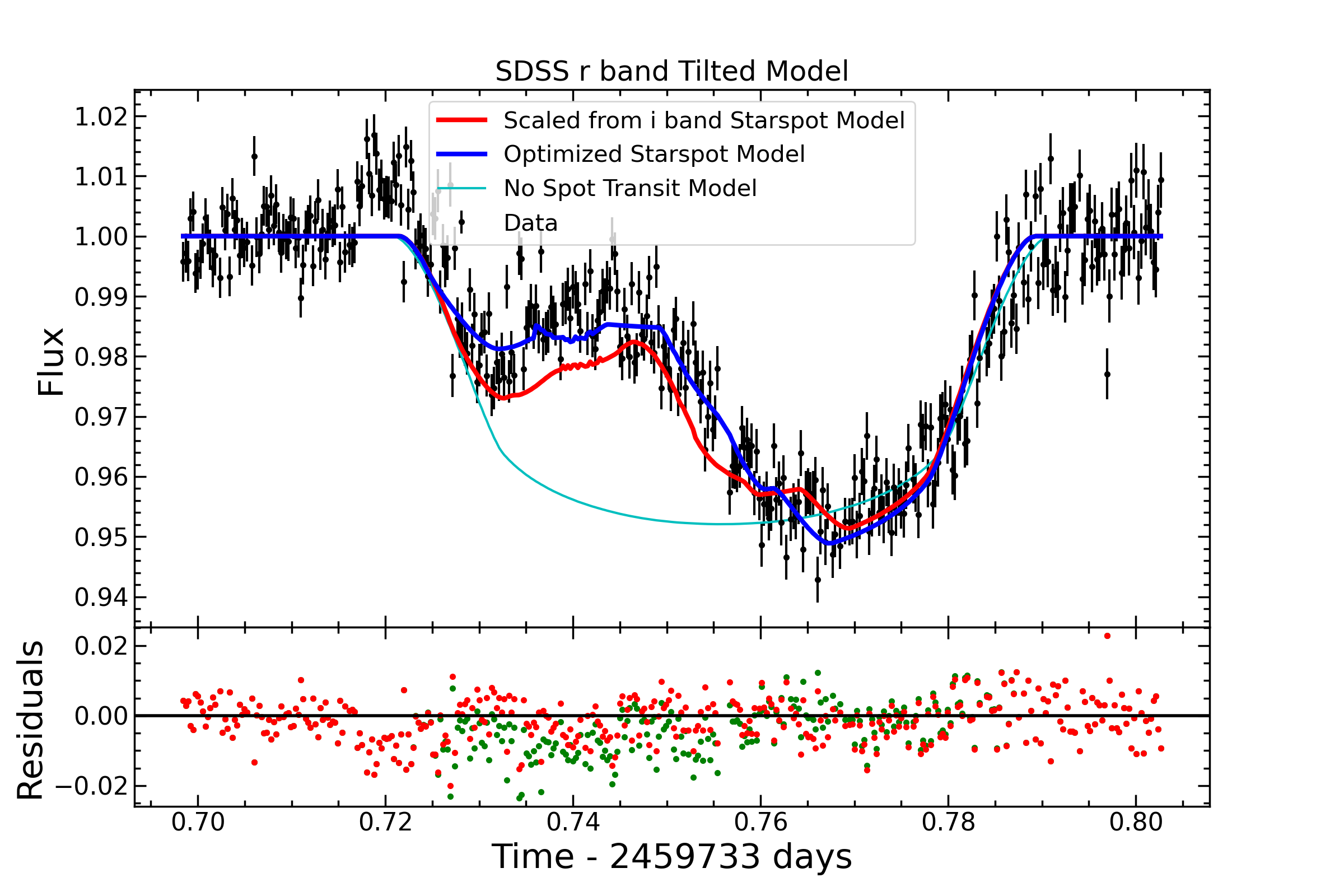} 
        \caption{\textbf{Top:} Projected star spots on TOI-3884's stellar surface for a polar star spot with a stellar spin axis tilt of $-65^{\circ}$ and $\lambda = 75^{\circ}$ using the APO SDSS \(r^\prime\) filter transit observed on 2022 June 03 assuming a spot temperature of 2900 K and a photospheric temperature of 3200 K (spot contrast of 0.7 for r$^\prime$ band). The large blue spot in the middle has a relative radius R$_{\rm spot}/$R$_s$  = 0.31 with the two smaller blue spots having radii R$_{\rm spot}/$R$_s$  = 0.22 and 0.11 respectively with the red starspots corresponding to the same starspots shown in Figure \ref{fig:sphere_polar}. The fractional spotted area in the transit chord for the blue stellar surface features assuming there are no spots anywhere else on the star is 4\%. \textbf{Middle:} Best fit starspot model for the oblique (not aligned) system shown in blue line compared to the no starspot transit model in cyan with the APO \(r^\prime\) band data as black points with error bars. The red line is the STSP model created by extending the SDSS \(i^\prime\) polar spot model to the SDSS \(r^\prime\) contrast. \textbf{Bottom:} Residuals from the best-fit star spot model (blue points) and scaled from SDSS \(i^\prime\) polar spot model (red points).}\label{fig:sphere_r}
    \end{minipage}
    
\end{figure}

For the APO SDSS r$^\prime$ transit, we chose to assume the same number of spots and spot temperature while allowing the location and spot radii to vary.  We discover that the polar spot remains approximately the same radius but shifts slightly while the other two spots slightly increase in area (blue spots in Figure~\ref{fig:sphere_r}). Thus, while the general location of the features near the pole are consistent across six months, the individual star spots are most likely evolving from one transit to the next. This tentative evidence for small-scale spot changes suggests caution when directly comparing transits observed at different times. If we instead allow the spot temperature to change rather than the spot location and radii, we find that the required contrast to fit the APO SDSS r$^\prime$ transit is nearly perfectly dark (c = 0.90) if we assume the same spot configuration as found in Figure \ref{fig:sphere_polar}. Since the spot contrast required to fit the SDSS r$^\prime$ transit is unreasonably dark, it is more likely there is small-scale spot evolution between transits.


\section{Discussion}\label{sec:discussion}

\subsection{Comparison to Previous Work}\label{subsec:comparison}
While our qualitative conclusions generally agree with those from \citet{almenara.toi3884}, there are notable exceptions.

\textit{Stellar Rotation Period:} We determine a stellar \vsini~of 3.59 $\pm$ 0.92 km/s for TOI-3884 in contrast to \citet{almenara.toi3884} who find a \vsini~of 1.1 km/s. We arrive at the more rapid rotational value both via \texttt{HPF-SpecMatch} and CCF methods. We also attempted to duplicate their method by using a template star near-identical to their comparison star, LHS 1140\footnote{LHS 1140's declination is inaccessible to the HET.} which is slightly cooler and lower mass than TOI-3884. However, we still obtain a \vsini~of 3.5 km/s. Using the LAMOST-derived H$\alpha$ EW, models from \citet{newton.halpha.relation} constrain TOI-3884's rotation period $<$ 10 days -- faster than a \vsini~of 1.1 km/s. 

\citet{almenara.toi3884} do note significant variation in their CCF profiles between their two spectra, likely a result of the large spot. In turn, we do not observe significant variation in either the differential line width (dLW) or chromatic index (CRX) across the 34 unbinned spectra. We attribute this to HPF's near-infrared wavelength range as well as HPF's lower resolution of R$\sim$53,000 (compared to ESPRESSO's bluer wavelength coverage and higher resolution). Thus it is also possible the discrepancy between reported \vsini~ is due to stellar activity -- which HPF is less affected by.

\textit{Planetary Mass:} \citet{almenara.toi3884} derive a planetary mass of 16.5$^{+3.5}_{-1.8}$ M$_\oplus$ using two RV ESPRESSO points. This is a 2.2$\sigma$ discrepancy from our higher mass measured with HPF. We checked for correlations between the HPF RVs and the dLW and CRX. We found no statistically significant correlation with a $\rho$ of -0.191 (p-value: 0.282) and 0.171 (p-value: 0.334) for the respective activity indicators. It is therefore unlikely that activity alone is amplifying the periodic planetary signal observed by HPF. We attempted to jointly fit both the HPF and ESPRESSO RV points, however, our model requires a 14 m/s jitter term added to the ESPRESSO RVs. ESPRESSO's bandpass of 380 -- 780 nm is more susceptible to stellar activity \citep{reiners.nIR.RV.2010}. \texttt{SOAPv2} approximates our pole-spot model should introduce a $\sim$10 m/s signal into optical ESPRESSO RVs -- similar to the required jitter of our model. With 17 near-IR HPF RVs, we robustly measure a 4$\sigma$ planetary mass.

\textit{Planetary Radius and Transit Depth Chromaticity:} We measure a larger planet of 6.43 $\pm$ 0.20 R$_\oplus$ compared to \citet{almenara.toi3884} who report two different radii: 6.31 $\pm$ 0.28 R$_\oplus$ from GP fitting and 6.00 $\pm$ 0.18 R$_{\oplus}$ from \texttt{starry}. We test the impact of our transit mask on the best-fit transit depth and planetary radius derived from the joint fit. To accomplish this we fit a transit model to the 2022 April 05 APO observation without any mask deriving a planetary radius of 6.05 $\pm$ 0.14 R$_\oplus$ -- comparable to \citet{almenara.toi3884} but a poor fit to the APO light curve. We then created ten random spot masks of various positions and sizes though required the mask to include the most extreme spot crossing event (between 30 minutes pre-transit to 7 minutes post-transit). A transit model was again fit to these light curves where we derived planetary radii spanning 6.39 to 6.61 R$_\oplus$, well within 1$\sigma$ of our reported planetary radius as well as in good agreement with \citet{almenara.toi3884} GP-derived radius for TOI-3884b. Therefore the discrepancy between the two reported radii is not dependent on our mask selection alone.

When further investigating this discrepancy, we discovered that their transits \citep[see Figure 1 in][]{almenara.toi3884} demonstrate significant chromatic variability even outside of the modeled pole-spot. Chromatic transits can either be explained via a background eclipsing binary \citep{wang.chromatic.photometry}, or unocculted stellar activity \citep{rackham_tlse_2018}. As we rule-out nearby companions, we explore the impact of stellar activity on our transit depth. 

We fit a transit model to the two masked APO SDSS i$^\prime$ and SDSS r$^\prime$ transits holding a/R$_s$, impact parameter, transit ephemeris, and transit depth (R$_{p}$/R$_{s}$)$^{2}$ constant across both transits. We check for chromaticity in the masked APO transit depths between the SDSS i$^\prime$ and SDSS r$^\prime$ observations which could indicate a contaminating background source. Using the two masked transits we fit a transit using \texttt{exoplanet} \citep{foreman-mackey_exoplanet_2021} holding a/R$_s$, impact parameter, transit ephemeris, and a dilution term multiplied to the SDSS r$^\prime$ transit depth. Assuming no chromaticity, the dilution term is equal to one. We determine a dilution term of 1.01 $\pm$ 0.02. There is no significant transit depth difference between these two wavelength bandpasses. However, both these depths are slightly shallower than the bluer g' transit in \citet{almenara.toi3884} and deeper than the IR ExTRA observation. Differing spot contrasts compared to the hotter photosphere creates deeper transits at bluer wavelengths \citep{rackham_tlse_2018}. We approximate by eye the depths of their individual transits. We fit both our and their wavelength-dependent depths using a simple unocculted star spot model from \citet{rackham_tlse_2018}:
\begin{equation}
D_\textrm{obs} = \frac{D}{1-f{_\textrm{sp}}(1-\frac{F_{\lambda,sp}}{F{_\lambda,ph}})}
\end{equation}

where D$_\textrm{obs}$ is the wavelength-dependent observed transit depth, D is the true transit depth, f$_\textrm{sp}$ is the spot coverage fraction, and F$_\lambda$ is the wavelength-dependent flux of the spot (sp) and photosphere (ph) respectively. The unocculted spot model fits the four transit depths assuming a spot temperature of 2900 K, total unocculted spot coverage of 16\%, and a true planet radius of $\sim$6.2 R$_\oplus$. We do not include uncertainties in these numbers as the \citet{almenara.toi3884} transit depths and uncertainties are relied on by-eye approximations. However, this model demonstrates that the chromatic transit depth is explained by unocculted stellar activity which slightly impacts the measured radius of the planet ($\sim$1$\sigma$ discrepancy from our radius).

\textit{Stellar Inclination and Spot Properties:} We fit a spot model based on the values reported in \citet{almenara.toi3884} to the SDSS i$^\prime$ transit, determining a best-fit $\chi^2_r$ of 6.09. Assuming the spot is evolving over time, it is possible that the spot evolved between the two observations. However, their spot model generates significant ($\sim$1\%) out of transit variability that is not observed in the TESS nor ground-based photometry. Accounting for the lack of baseline variability enabled us to constrain the stellar inclination to 25 $\pm$ 5$^\circ$ which in turn impacted our overall best-fit spot model.

\subsection{Comparison of TOI-3884b in M Dwarf Planetary Parameter-Space}\label{sec:parameterspace}

\begin{figure*}[!t]
\fig{MR_3884_annotated.PNG}{0.45\textwidth}
{\small a) Planet radius as a function of mass}    \label{fig:RadiusMass}
\fig{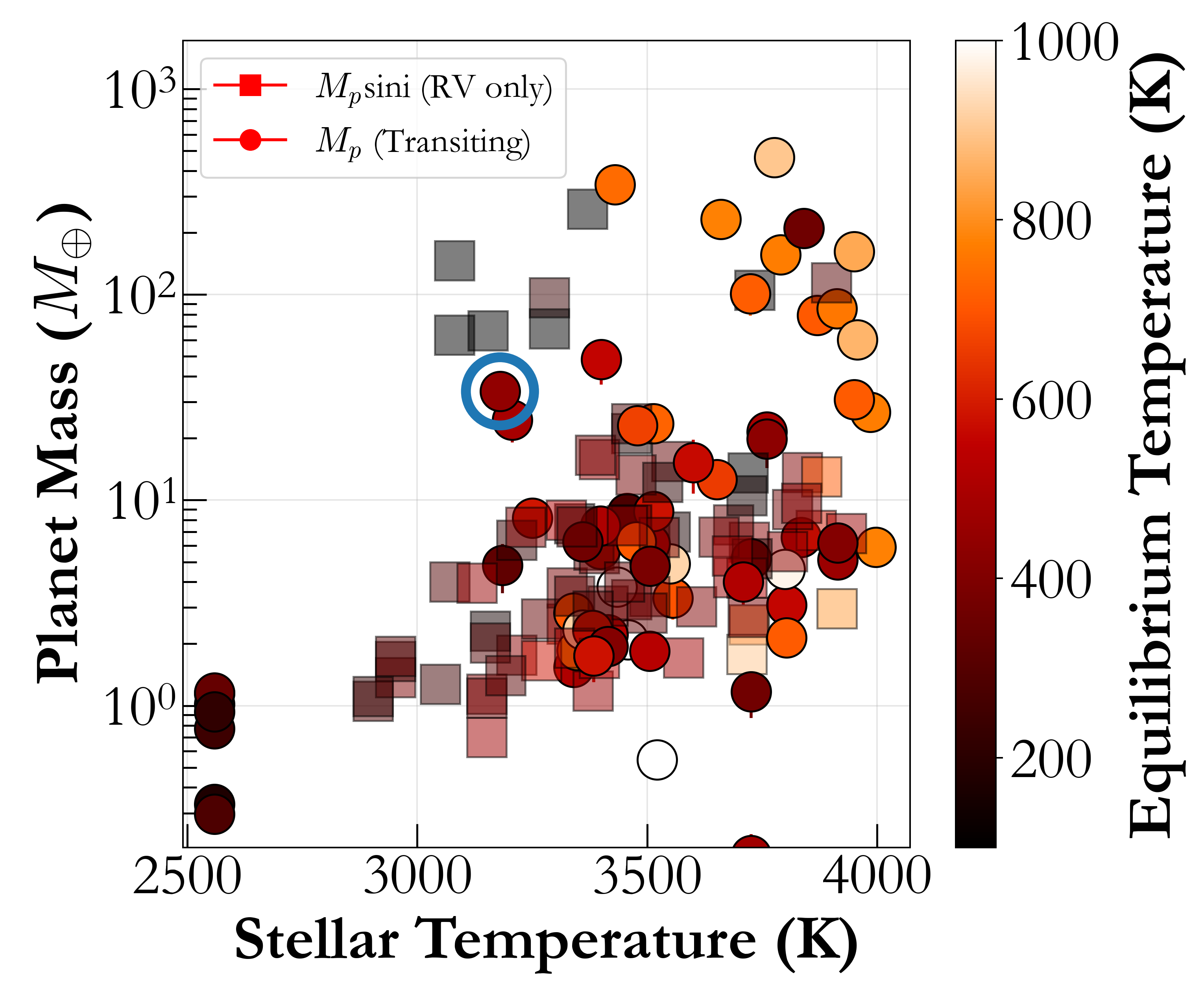}{0.45\textwidth}
{\small b) Planet mass as a function of \teff{}} \label{fig:RadiusTeff} \vspace{0.2 cm}
\caption{\small Sample of transiting M dwarf planets that have precise mass measurements ($> 3 \sigma$). \textbf{a)} Mass-Radius plane showing the small sample ($\sim 15$) of giant planets ($R_p > 4$ \earthradius{}) orbiting M dwarfs (\teff{} $<$ 4000 K), color coded by \teff{}. \textbf{b)} The masses for all M dwarf planets as a function of \teff{}, showing how TOI-3884b stands out in terms of its stellar host. Transiting planets are shown as circles, whereas RV only (m sin$i$) detections are in squares. The clump of planets at $\sim 2600$ K represent the TRAPPIST-1 system \citep{grimm_nature_2018}.}\label{fig:ParameterSpace}
\end{figure*}

Super-Neptunes (4 \earthradius{} $< R_p < 8$ \earthradius{}) represent a transitional population of planets between the rocky terrestrial planets and Jovian gas giants. TOI-3884b adds to the growing sample of well-characterized (with precise 3$\sigma$ masses and radii) super-Neptunes orbiting M dwarfs. In particular, \autoref{fig:ParameterSpace}a shows TOI-3884b's position in a planetary mass-radius plane with respect to other M dwarfs (\teff $< 4000$ K) planets with known ($> 3\sigma$) masses and radii. Figure \ref{fig:ParameterSpace}b plots the same sample as a function of stellar effective temperature.  


The formation of Jovian planets around M dwarfs should be inhibited by the longer orbital time scales with respect to the disk lifetimes \citep{laughlin.core.accretion, ida_toward_2005}. This is corroborated by empirical data from RV surveys \citep{endl_exploring_2006, johnson_giant_2010, maldonado_connecting_2019, sabotta_carmenes_2021}. However, \cite{laughlin.core.accretion} predict that M dwarfs should host an abundance of Neptunes that fail to accrete a massive enough core \citep[$\sim 10$ \earthmass; {}][]{pollack_formation_1996} in a timely manner to initiate runaway gaseous accretion. Using models from \citet{fortney.core.2007} and propagating the uncertainties in planetary parameters using the Monte-Carlo method, we predict TOI-3884b's core mass to be about 21 $\pm$ 4 \earthmass. It should therefore have experienced some runaway gaseous accretion. The fact that TOI-3884b did not accrete a Jovian-mass atmosphere suggests that its core was slow to form, or there was a lack of nearby gas/material for rapid accretion or both.  

\subsection{Atmosphere of TOI-3884b}\label{subsec:atmosphere}

\begin{figure}
    \centering
    \includegraphics[width=\columnwidth]{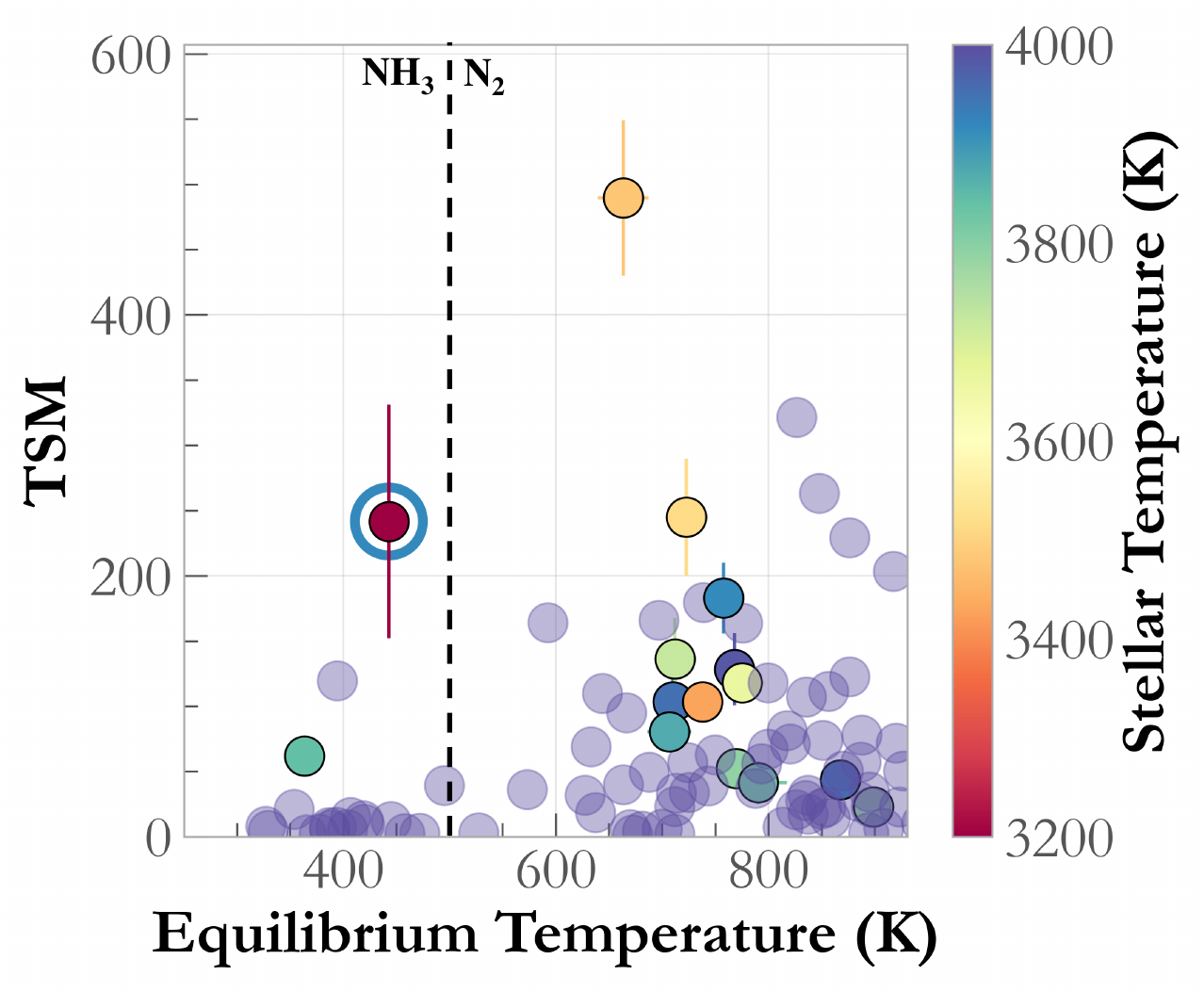}
    \caption{Transmission Spectroscopy Metric (TSM) as a function of planetary equilibrium temperature for all planets with a known ($>$3$\sigma$) mass and T$\mathrm{eq}$ cooler than 1000 K. Points are colored based on their host star's effective temperature with planets around M dwarfs denoted with solid coloring. The approximate temperature when ammonia appears in a planet's atmosphere (assuming equilibrium chemistry) is denoted with the black dashed line. TOI-3884b (blue circle) possesses one of the highest TSMs of any non-Hot-Jupiter and the highest TSM for planets $<$ 500 K.}
    \label{fig:TSM}
\end{figure}

While the spot portion of the transit may complicate the analysis, TOI-3884b has the highest transmission spectroscopic metric \citep[TSM;][]{kempton_framework_2018} of any known planet with an equilibrium temperature $<$ 500 K (TSM: 230 -- Figure~\ref{fig:TSM}). Owing to its bright host star and large transit depth, this planet also has one of the highest metrics of any known non-Hot Jupiter planet. At 430 K, TOI-3884b's atmosphere likely contains methane as the carbon-dominant molecule along with water and some ammonia -- assuming equilibrium chemistry \citep{zahnle_bdchemistry_2012,kempton_exo-transmit_2017,fortney_beyond_2020,hu_cold_chemistry_2021}. Deriving the overall abundances of these molecules would provide an approximate C/N/O ratio, a useful measurement for constraining where this planet originally formed in its disk \citep{oberg_nitrogen_2019,turrini_cno_ratios_2021,hobbs_cno_ratio_2022}. \citet{oberg_co_ratios_2011} demonstrate the connection between C/O ratios and various molecular snow-lines. \citet{dash_cno_ratios_2022} expand on this study by noting that nitrogen provides information surrounding the disk's overall metallicity, as it is unaffected by the condensation of molecules such as water, carbon dioxide, and methane. Only the ammonia snow-line at $\sim$100 K and N$_{2}$ snow-line at $\sim$78 K significantly affects its overall ratio in the disk. TOI-3884b is therefore an extremely promising target to observationally test the link between nitrogen abundance and formation location.

Due to the combination of its cool equilibrium temperature along with experiencing UV-radiation from its active M dwarf host, TOI-3884b's atmosphere is likely comprised of photochemically created hazes such as tholins \citep[e.g.,][]{morley_gj1214b_haze_2013} or even soot \citep{gao_sulfur_hazes_2017}. Photochemically created hazes \citep[e.g.,][]{tsai.jwst.photochemistry}, and aerosols in general, are common in exoplanetary atmospheres with several studies linking their presence to temperature \citep{crossfield_kreidberg_2017,dymont_cleaning_2021,yu_hazes_trends_2021}. Assuming the trend highlighted in \citet{yu_hazes_trends_2021} holds, we would expect TOI-3884b to possess a fairly hazy transmission spectrum in the near-infrared. However, \citet{kawashima_haze_translucent_2019} show that hazes should become translucent at longer wavelengths assuming the overall particle sizes are small. Thus, extending out to $>$3 $\mu$m should enable atmospheric characterization of TOI-3884b regardless of its expected hazy atmosphere.

\begin{figure}
    \centering
    \includegraphics[width=\columnwidth]{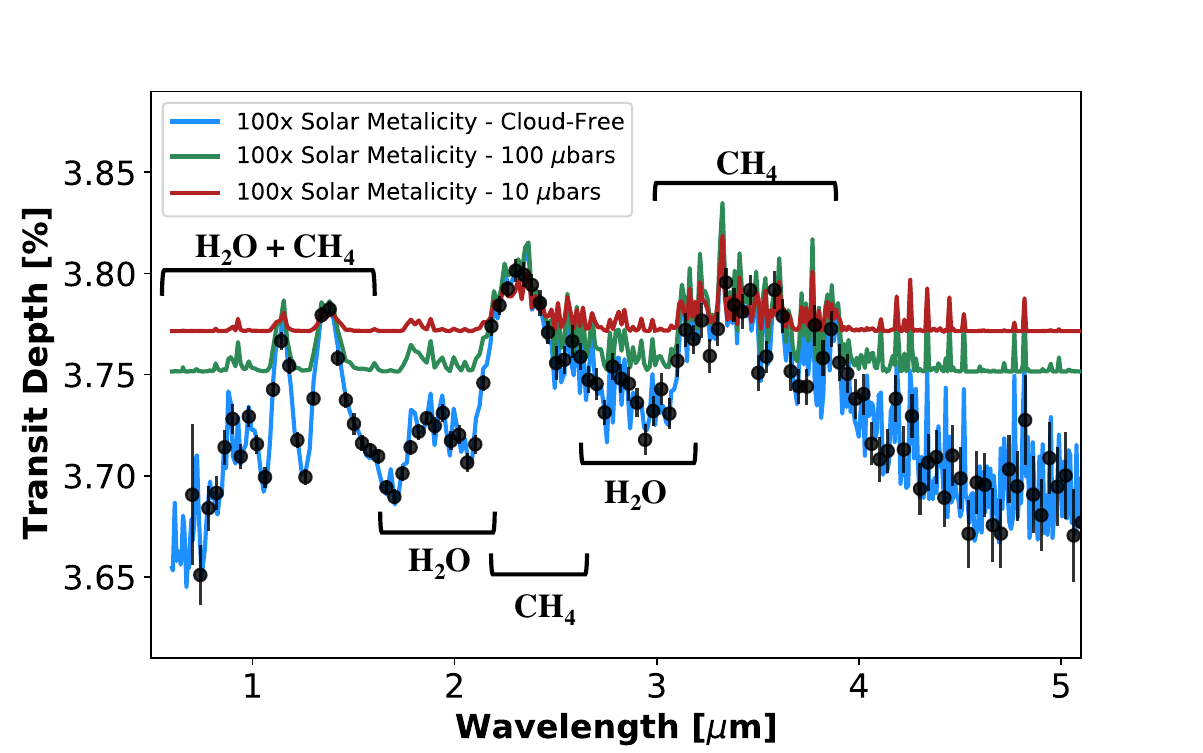}
    \caption{Transmission spectra generated with \texttt{ExoTransmit} for a 100$\times$ Solar metallicity atmosphere in chemical equilibrium with gray-absorber aerosol layer at 10 and 100 $\mu$m. Simulated data is created using \texttt{PandExo} for two transits of JWST-NIRSpec and based on the cloud-free model. The two dominant absorbers, water and methane, are labeled for reference though other molecules including ammonia are also included in the models.}
    \label{fig:spectrum}
\end{figure}

We generate the expected transmission spectrum of TOI-3884b using \texttt{ExoTransmit} \citep{kempton_exo-transmit_2017} assuming a 100$\times$ Solar metallicity atmosphere with no aerosols, aerosols at pressures of 100 $\mu$bars and aerosols at pressures of 10$\mu$bars (Figure~\ref{fig:spectrum}). For these simulations we assume a gray-opacity aerosol layer which is wavelength independent. Using the cloud-free model, we used PandExo \citep{batalha_pandexo_2017} to simulate two transit observations with \textit{JWST} NIRSpec-Prism. We find that we should easily retrieve methane and water in both the cloud-free and 100 $\mu$bars cases. At 10 $\mu$bars we will still observe methane absorption features with tentative detection of water. It should be noted however, that these simulations assume a typical transit shape which allows for easily derived uncontaminated transit depths. Assuming the bump-feature in TOI-3884b's transit remains long-lived, it is possible that the uncertainties presented in Figure~\ref{fig:spectrum}, are underestimated and stellar contamination will also need to be included in modeling the observed transmission spectrum.

TOI-3884b presents a second unique opportunity: the impact of starspots on the transmission spectrum of a planet. Stellar activity due to an inhomogeneous photosphere may introduce spurious features into a planet's transmission spectrum \citep[e.g.,][]{rackham_tlse_2018,barclay_k218b_2021}. This is of particular concern with M-dwarfs, whose spots are cool enough to host their own water absorption features \citep{Jones_watervapor_1995}. \citet{rackham_spotcontamination_exoplanetexploration} emphasizes the need for future in-depth studies into untangling star and planetary spectra especially as we begin to probe the atmospheres of terrestrial worlds with JWST and future instruments. With half of its transit covered by a spot and the other over the photosphere, comparing the resulting transmission spectrum from the first half of the transit to the second may yield an unprecedented probe into the effects of cooler spots.

\subsection{Orbital Alignment of TOI-3884b}

Assuming the pole-spot hypothesis, TOI-3884b possesses a misaligned orbit with an obliquity of 75 $\pm$ 10$^{\circ}$. TOI-3884b therefore joins the growing population of misaligned warm-Neptunes (R $>$4 \earthradius), which includes the two M dwarf Neptunes: GJ 3470b \citep{stefansson_gj3470orbit_2022} and GJ 436b \citep{bourrier_gj436_misaligned_2018}. 

Neither Gaia nor the HPF RV residuals detect evidence of any outer massive companion in the TOI-3884 which could have been responsible for TOI-3884b's misaligned orbit \citep{petrovich_planetobliquity_2020}. Of the four misaligned Neptunes orbiting K and M dwarfs, two \citep[HAT-P-11b and WASP-107b][]{yee_hatp11_2018,piaulet_wasp107b_2021} have a confirmed outer companion while \citet{stefansson_gj3470orbit_2022} does not exclude the existence of an outer planet in the GJ 3470 system. Giants around M dwarfs are uncommon; it is unlikely that TOI-3884 hosts an additional gas giant responsible for the misalignment of TOI-3884b. However, our RV observations are limited to $<$ 6 months. Continued radial velocity monitoring is required to detect longer period massive planets in this system. 

\section{Conclusion}\label{sec:conclusion}

We confirm the planetary nature of TOI-3884b, a super-Neptune crossing a persistent spot during transit. This spot-crossing event is chromatic, and we conclude this bump is created by a large star spot which appears at the same location in every transit spanning over a year of monitoring. We present two hypotheses: 1) TOI-3884's rotation is exactly equal to its planet's orbital period of 4.56 d, or 2) TOI-3884 rotational axis is tilted along our line of sight and TOI-3884b crosses a polar spot. Given the lack of significant photometric or spectroscopic variability in the RVs, \tess~light curves, and ground-based monitoring spanning over six months, we strongly prefer the second pole-spot hypothesis. In this scenario, TOI-3884's spin-axis is inclined along our line-of-sight. TOI-3884b therefore possesses a misaligned orbit that is nearly polar to its star. TOI-3884b joins the population of misaligned warm-Neptunes around low-mass stars \citep{albrecht_obliquities_2022}. 

We also discover signs of spot evolution between the different transits. While the in-transit bump appears at a similar position, its overall structure changes on measurable timescales. The TOI-3884 system presents a rare opportunity to monitor pole-spot evolution on an active mid-M dwarf. 


\section{Acknowledgements}

We thank the anonymous referee for their thoughtful suggestions which greatly improved this work. We also thank Will Waalkes and Michael Gully-Santiago for useful discussions. 
The Center for Exoplanets and Habitable Worlds is supported by the Pennsylvania State University and the Eberly College of Science.
The computations for this research were performed on the Pennsylvania State University's Institute for Computational and Data Sciences' Roar supercomputer, including the CyberLAMP cluster supported by NSF grant MRI-1626251. This content is solely the responsibility of the authors and does not necessarily represent the views of the Institute for Computational and Data Sciences.

The Pennsylvania State University campuses are located on the original homelands of the Erie, Haudenosaunee (Seneca, Cayuga, Onondaga, Oneida, Mohawk, and Tuscarora), Lenape (Delaware Nation, Delaware Tribe, Stockbridge-Munsee), Shawnee (Absentee, Eastern, and Oklahoma), Susquehannock, and Wahzhazhe (Osage) Nations.  As a land grant institution, we acknowledge and honor the traditional caretakers of these lands and strive to understand and model their responsible stewardship. We also acknowledge the longer history of these lands and our place in that history.

We acknowledge support from NSF grants AST 1006676, AST 1126413, AST 1310875, AST 1310885, AST 2009554, AST 2009889, AST 2108512, AST 2108801 and the NASA Astrobiology Institute (NNA09DA76A) in our pursuit of precision RVs in the near-infrared. We acknowledge support from the Heising-Simons Foundation via grant 2017-0494.

We acknowledge support from NSF grants AST 1907622, AST 1909506, AST 1909682, AST 1910954 and the Research Corporation in connection with precision diffuser-assisted photometry.

CIC acknowledges support by NASA Headquarters through an appointment to the NASA Postdoctoral Program at the Goddard Space Flight Center, administered by USRA through a contract with NASA.

GS acknowledges support provided by NASA through the NASA Hubble Fellowship grant HST-HF2-51519.001-A awarded by the Space Telescope Science Institute, which is operated by the Association of Universities for Research in Astronomy, Inc., for NASA, under contract NAS5-26555.

JW acknowledges assistance from NSF grant AST 1907622.
WDC acknowledges support from NSF grant 2108801. 
This work is Contribution 0046 from the Center for Planetary Systems Habitability at the University of Texas at Austin.  These results are based on observations obtained with HPF on the HET. The HET is a joint project of the University of Texas at Austin, the Pennsylvania State University, Ludwig-Maximilians-Universit\"at M\"unchen, and Georg-August Universit\"at Gottingen. The HET is named in honor of its principal benefactors, William P. Hobby and Robert E. Eberly. The HET collaboration acknowledges the support and resources from the Texas Advanced Computing Center. We are grateful to the HET Resident Astronomers and Telescope Operators for their valuable assistance in gathering our HPF data.

WIYN is a joint facility of the University of Wisconsin-Madison, Indiana University, NSF's NOIRLab, the Pennsylvania State University, Purdue University, University of California-Irvine, and the University of Missouri. 

Based on observations at Kitt Peak National Observatory, NSF’s NOIRLab, managed by the Association of Universities for Research in Astronomy (AURA) under a cooperative agreement with the National Science Foundation. The authors are honored to be permitted to conduct astronomical research on Iolkam Du’ag (Kitt Peak), a mountain with particular significance to the Tohono O’odham. 
Deepest gratitude to Zade Arnold, Joe Davis, Michelle Edwards, John Ehret, Tina Juan, Brian Pisarek, Aaron Rowe, Fred Wortman, the Eastern Area Incident Management Team, and all of the firefighters and air support crew who fought the recent Contreras fire. Against great odds, you saved Kitt Peak National Observatory.

Some of results are based on observations obtained with the Apache Point Observatory 3.5 m telescope, which is owned and operated by the Astrophysical Research Consortium. We wish to thank the APO 3.5 m telescope operators in their assistance in obtaining these data.

Some of the observations in this paper made use of the NN-EXPLORE Exoplanet and Stellar Speckle Imager (NESSI). NESSI was funded by the NASA Exoplanet Exploration Program and the NASA Ames Research Center. NESSI was built at the Ames Research Center by Steve B. Howell, Nic Scott, Elliott P. Horch, and Emmett Quigley.

Some of the data presented in this paper were obtained from MAST at STScI. Support for MAST for non-HST data is provided by the NASA Office of Space Science via grant NNX09AF08G and by other grants and contracts.

This work includes data collected by the TESS mission, which are publicly available from MAST. Funding for the TESS mission is provided by the NASA Science Mission directorate. 

The TESS data presented in this paper were obtained from the Mikulski Archive for Space Telescopes (MAST) at the Space Telescope Science Institute. The specific observations analyzed can be accessed via \dataset[DOI: 10.17909/t9-r086-e880]{https://doi.org/10.17909/t9-r086-e880} and \dataset[DOI: 10.17909/t9-wpz1-8s54]{https://doi.org/110.17909/t9-wpz1-8s54}.

This research made use of the (i) NASA Exoplanet Archive, which is operated by Caltech, under contract with NASA under the Exoplanet Exploration Program, (ii) SIMBAD database, operated at CDS, Strasbourg, France, (iii) NASA's Astrophysics Data System Bibliographic Services, (iv) NASA/IPAC Infrared Science Archive, which is funded by NASA and operated by the California Institute of Technology, and (v) data from 2MASS, a joint project of the University of Massachusetts and IPAC at Caltech, funded by NASA and the NSF.

This work has made use of data from the European Space Agency (ESA) mission Gaia (\url{https://www.cosmos.esa.int/gaia}), processed by the Gaia Data Processing and Analysis Consortium (DPAC, \url{https://www.cosmos.esa.int/web/gaia/dpac/consortium}). Funding for the DPAC has been provided by national institutions, in particular the institutions participating in the Gaia Multilateral Agreement.

Some of the observations in this paper made use of the Guoshoujing Telescope (LAMOST), a National Major Scientific Project built by the Chinese Academy of Sciences. Funding for the project has been provided by the National Development and Reform Commission. LAMOST is operated and managed by the National Astronomical Observatories, Chinese Academy of Sciences.

Some of the observations in this paper were obtained with the Samuel Oschin Telescope 48-inch and the 60-inch Telescope at the Palomar Observatory as part of the ZTF project. ZTF is supported by the NSF under Grant No. AST-2034437 and a collaboration including Caltech, IPAC, the Weizmann Institute for Science, the Oskar Klein Center at Stockholm University, the University of Maryland, Deutsches Elektronen-Synchrotron and Humboldt University, the TANGO Consortium of Taiwan, the University of Wisconsin at Milwaukee, Trinity College Dublin, Lawrence Livermore National Laboratories, and IN2P3, France. Operations are conducted by COO, IPAC, and UW.

\facilities{ARC (ARCTIC), Exoplanet Archive, Gaia, HET (HPF), LAMOST, MAST, PO:1.2m (ZTF), PO:1.5m (ZTF), TESS, WIYN (NESSI)} 
\software{
AstroImageJ \citep{collins_astroimagej}, 
\texttt{astropy} \citep{astropy_collaboration_astropy_2018},
\texttt{barycorrpy} \citep{kanodia_python_2018}, 
\texttt{EXOFASTv2} \citep{eastman_exofastv2_2019},
\texttt{exoplanet}\citep{foreman-mackey_exoplanet_2021}
\texttt{HPF-SpecMatch} (S. Jones et al. 2022),
\texttt{isochrone} \citep{isochrones},
\texttt{lightkurve} \citep{lightkurve_collaboration_lightkurve_2018},
\texttt{PyMC3} \citep{salvatier_probabilistic_2016},
\texttt{pyHammer}\citep{roulston_pyhammer_2020},
\texttt{STSP},\citep{morris2017,schutte_starspot_2022}
}

\bibliography{references,references2}

\end{document}